\documentclass[twocolumn,showpacs,amsmath,amssymb,amsfonts]{revtex4}

\usepackage{graphicx}
\usepackage{dcolumn}
\usepackage{bm}
\usepackage{latexsym,color}

\begin{document}

\title{Modal Approach to Casimir Forces in Periodic Structures}

\author{P. S. Davids$^{1}$}
\author{F. Intravaia$^2$}
\author{F. S. S. Rosa$^2$}
\author{D.A.R. Dalvit$^2$}
\affiliation{%
$^{1}$Applied Photonics and Microsystems, Sandia National Laboratories, Albuquerque, NM 87185, USA\\
$^2$Theoretical Division, MS B213, Los Alamos National Laboratory, Los Alamos, NM 87545, USA}%

\begin{abstract} 
We present a modal approach to calculate finite temperature Casimir interactions between two periodically modulated surfaces. The scattering formula is used and the reflection matrices of the patterned surfaces are calculated decomposing the electromagnetic field into the natural modes of the structures. The Casimir force gradient from a deeply etched silicon grating is evaluated using the modal approach and compared to experiment for validation.  The Casimir force from a two dimensional periodic structure is computed and deviations from the proximity force approximation examined.
\end{abstract}


\maketitle


\section{Introduction}

The dynamics of a classical or a quantum field drastically depends on the external boundary conditions imposed on it. These boundary conditions lead to a modification of its power spectrum with consequences on measurable quantities like the radiation pressure. One of the most notable examples of this kind  of phenomena is the Casimir effect \cite{Casimir48}: in its original formulation the change in the spectral density of zero-point fluctuations of the quantum electromagnetic field induced by the presence of two perfectly reflecting parallel plates turns into a net radiation pressure that pushes one plate towards the other or, in other words, into an attractive force between the plates.  

Recently, thanks to technological advancements, we have witnessed an increased interest in Casimir interactions \cite{Bordag01,Milton04,Lamoreaux05,Lamoreaux97,Mohideen98,Chan01s,Bressi02,Decca03}. Indeed, the Casimir force offers new possibilities for nanotechnology, such as actuation in micro- and nanoelectromechanical systems (MEMS and NEMS) mediated by  the quantum vacuum. However, it also presents some challenges since the same force is generally recognized as one of the possible sources of stiction and consequently of malfunctioning of these devices.
Researchers have therefore started an intense theoretical \cite{Henkel05,Rosa08a,Rahi10a,Silveirinha10,Rodrigues07,Zhao09,McCauley10a,Levin10} and experimental \cite{Lee01,Munday09} program in order put some theoretical constraints \cite{Henkel05,Rosa08a,Rahi10a,Silveirinha10} and to understand how to engineer the strength and possibly also the sign of the Casimir force - a repulsive Casimir force would provide an anti-``stiction" effect. Inevitably a lot of attention has been focused on the role of the boundary conditions and very recently on the interplay of material properties, temperature, and geometry.  

In this paper we present our results for the computation of finite temperature Casimir forces between periodic nanostructures using a modal approach. 
Our calculation is based on the scattering approach to the Casimir effect which gives the Casimir force between two objects starting from their scattering properties. In our case this reduces the problem to the calculation of the reflection matrices of the periodic nanostructures. We will pay particular attention to 2D lamellar gratings, 
which are periodic metallic and/or dielectric structures that consist of planar layers. Many complex 3D structures, such as photonic crystals and metamaterials, can be thought of as being constructed from individual  2D extruded metallic and dielectric strata \cite{Li93a}. 
The problem of the reflection of an electromagnetic field impinging on a periodic structure is a topic that has a huge literature, and there are a large variety of methods to efficiently compute the reflected and transmitted electromagnetic power 
from such a surface (for a review see, for example, \cite{Petit80,Neviere98}). Among the most famous methods one finds the differential approach \cite{Neviere71}, the integral 
approach \cite{Maystre72}, and the modal approach \cite{Li93a}. Our numerical results are based on   the last technique that, for its characteristic of simplicity, flexibility, and stability lends itself to be the most adequate to our purpose. (See \cite{Johnson11} for  a review of several numerical techniques specifically adapted to the computation of Casimir forces.)

While other more general  numerical scattering techniques exist, the modal expansion of the electromagnetic field provides insight into the anatomy of the Casimir force.  This microscopic analysis  provides an understanding of the nature of the electromagnetic fluctuations that give rise to the Casimir force and a means to modify their contribution to the Casimir force \cite{Intravaia05,Intravaia07,Intravaia09,Intravaia10,Haakh10}. These dominant eigenmode and frequency contributions to the Casimir force can be identified within the modal expansion and their influence under perturbation of the permittivity of the structure explored. Our modal expansion, based on a planewave expansion of the fields and a Fourier decomposition of the permittivity of the structure, is limited to periodic structures but is the natural choice to examine the Casimir force in  lamellar structures like photonic crystals and metamaterials. 

The paper is organized as follows. In Section \ref{Casimir} we briefly review the calculation of the Casimir interaction free energy and force at finite temperatures within the scattering approach, and we specialize the general formula to periodic structures. In Section \ref{Modal} we describe the modal approach and its application to generic (non-lamellar) periodic structures. Our finite-temperature numerical code is then benchmarked in Section \ref{Results} against experimental data for the Casimir interaction between a metallic sphere and a doped Si 1D lamellar grating \cite{Chan08}. Related zero-temperature computations using the differential approach have been performed in \cite{Lambrecht08a,Lambrecht09, Lambrecht11} for dielectric 1D lamellar gratings, and in \cite{Chiu10} for metallic 1D sinusoidal gratings.
We also compute Casimir forces in two-dimensional structures consisting of 2D array of silicon pillars, or a 2D array of square holes (the complementary structure to the array of pillars). In the last section we discuss our results and prospects for future studies.


\section{The Casimir free energy}
\label{Casimir}


\subsection{General framework}

In the last few years several authors have developed many powerful  (semi)analytic \cite{Gies03,Bulgac06, Emig06,Rahi09, Bordag06, Dalvit06, Lambrecht06, Kenneth08} 
and numerical \cite{Rodriguez07, Rodriguez09} approaches to calculate Casimir forces. They allow for the treatment of quite general geometric and material setups going 
beyond the simple plane-plane geometry of Casimir's seminal calculation.
For our purposes, the scattering approach \cite{Rahi09,Lambrecht06} is the most convenient formulation. The advantage of this technique is that it exclusively relies on  knowledge of the reflection properties of the objects seen as isolated scatterers for the electromagnetic radiation. Indeed, for linear magneto-dielectric media, i.e. media that are completely characterized by  permittivity $\overleftrightarrow{\epsilon}(\omega)$ and  permeability $\overleftrightarrow{\mu}(\omega)$ tensors, virtual and real photons are treated at the same level \cite{Dalibard82}, so that vacuum fluctuations are scattered in the same way as real fields.

The interaction free energy between two bodies in thermal equilibrium at a temperature $T$ is given by 
\begin{equation}
\mathcal{F}(a) = \frac{1} {\beta} {\sum^{\infty}_{l=0}}'  {\rm Tr}\log \left[\mathbf{ 1} - {\cal S}_1\cdot\mathcal{X}_{12}(a)\cdot{\cal S}_2\cdot \mathcal{X}_{21}(a) \right] ,
\label{casimir:energy}
\end{equation}
where $\beta = 1/ k_{\rm B} T$. ${\cal S}_i$ is the scattering operator characterizing the i-th object seen from the other object as if it were isolated. $X_{12}(a)$ and $X_{21}(a)$ are the so called translation operators \cite{Rahi09} and they carry no information about the objects but only depend on the distance $a$ between the origins of appropriately chosen coordinate systems in each body. All operators depend on $l$ through the Matsubara (imaginary) frequencies $\omega_l= i \xi_l = i 2\pi l/\hbar \beta$ (the prime in the summation indicates that  the $l=0$ term is weighted by $1/2$), and their detailed expressions depend on the functional basis we choose to describe the electromagnetic field. The symbol ``${\rm Tr}$'' indicates the trace over all spatial degrees of freedom, making the final result basis independent.


\begin{figure}
\centering\includegraphics[width=7.5cm]{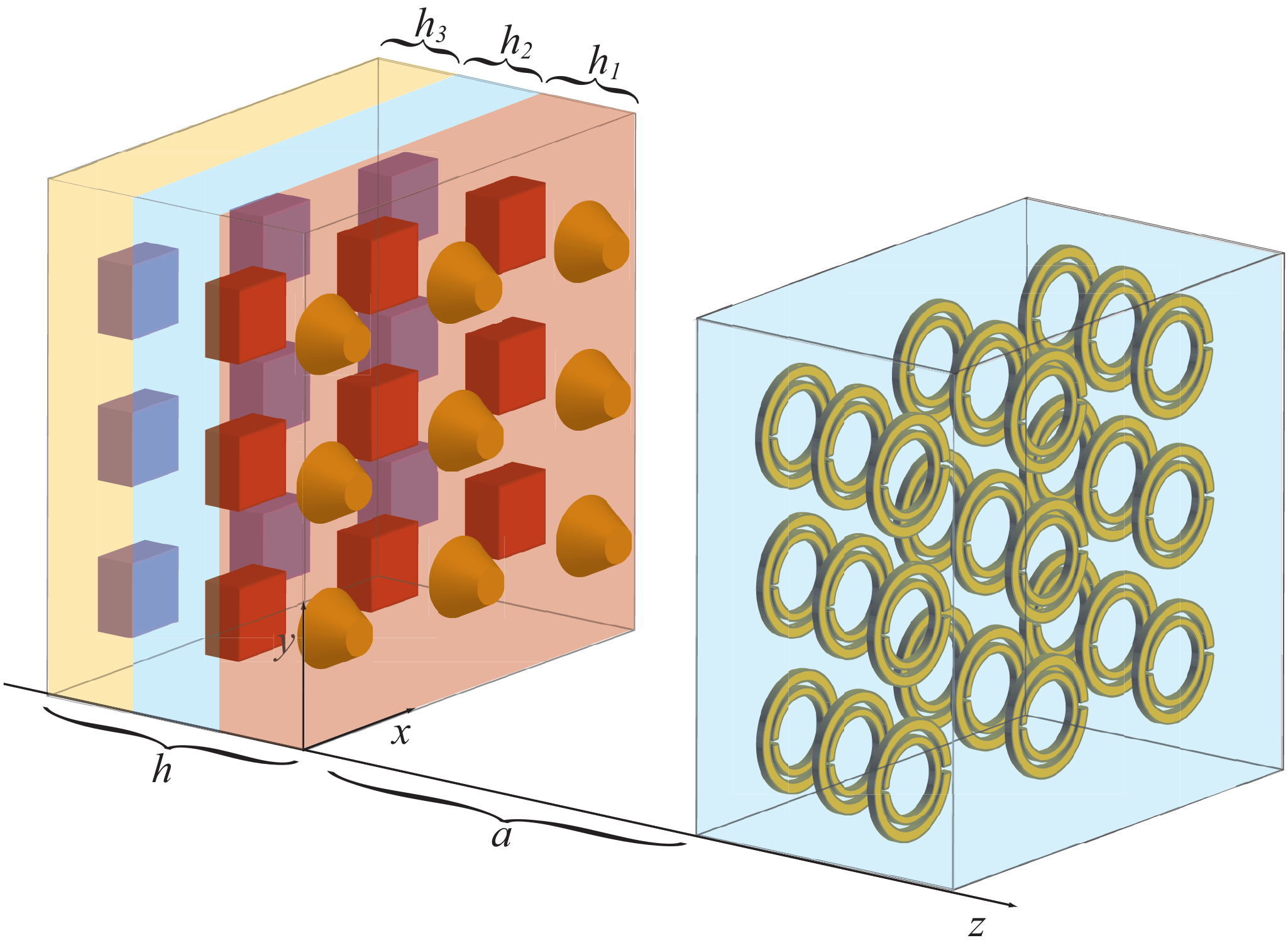}
\caption{Casimir set-up for two periodic structures (possibly multi-layered) parallel to each other and separated by a gap $a$. }
\label{fig:gratings}
\end{figure}


\subsection{Periodic systems}

The expression \eqref{casimir:energy}, being the infinite dimensional trace of a very involved operator, is in 
general an extremely complicated object. Most of the difficulty consists in finding the appropriate functional basis in which the operators in \eqref{casimir:energy} can be efficiently calculated and, when possible, reduce to  simple expressions. Most of the time the suitable 
basis for ${\cal S}_1$ is completely different from the one for ${\cal S}_2$. In such circumstances other matrices representing the change of bases have to be included in the representation of \eqref{casimir:energy} (see, for example \cite{Rahi09,Canaguier-Durand09}).

Fortunately, the symmetry properties of periodic systems suggest specific bases that allow for further manipulation and useful simplifications. This is the case of the simple setup where two gratings are facing each other shown if Fig. (\ref{fig:gratings}). The periods of the gratings are assumed to be the same for simplicity (for gratings with non-equal but commensurable periods the technique we are about to describe uses the
minimum common period for the two gratings).
The medium in between is assumed to be homogeneous  (it will become evident in the following that this condition can be easily relaxed). Both structures lie in the $(x,y)$ plane of a orthonormal cartesian coordinate system. The $z$ coordinate is called longitudinal because it is normal to both grating surfaces. The choice of a rectangular basis is not mandatory but dictated by symmetry, and it has been shown that sometimes the introduction of a nonrectangular coordinate system can improve the numerical convergence \cite{Li97}.

As usual for periodic system \cite{Ashcroft76,Kittel96},  it is convenient to introduce the transverse reciprocal space basis 
$\vert\mathbf{ k}_{||}, nm, \rightleftharpoons,\lambda \rangle $ associated with the reciprocal lattice described by the two dimensional vector
\begin{align}
\mathbf{K}_{||,n m} &\equiv \mathbf{k}_{||} + \frac{2\pi n}{L_x}\mathbf{\hat{x}} + \frac{2\pi m}{L_y}\mathbf{\hat{y}}\nonumber\\
&\equiv \alpha_{n}\mathbf{\hat{x}} + \beta_{m}\mathbf{\hat{y}} \quad n,m \in \mathbb{Z} ,
\end{align}
where we have implicitly defined
\begin{gather}
\mathbf{k}_{||}=k_{x}\mathbf{\hat{x}}+k_{y}\mathbf{\hat{y}}, \\
\alpha_{n}=k_{x}+ \frac{2\pi n}{L_x}\quad\text{and}\quad\beta_{m}=k_{y}+ \frac{2\pi m}{L_y} .
\end{gather}
$L_x$ and $L_y$ are the periods of the grating along the $\mathbf{\hat{x}}$ and $\mathbf{\hat{y}}$ directions, respectively.
The polarization is denoted by $\lambda$, and is chosen to be one of the orthogonal polarization,  $s$ or $p$, where $s$-polarized fields have the electric field along the
direction $\mathbf{\hat{e}}_s =  \mathbf{K}_{||,nm} \times  \mathbf{\hat{z}} / | \mathbf{K}_{||,nm} \times  \mathbf{\hat{z}}|$, and
$p$-polarized fields have the electric field along the direction
$\mathbf{\hat{e}}_p =  \mathbf{K}_{||,nm} \times  \mathbf{\hat{e}}_s / | \mathbf{K}_{||,nm} \times  \mathbf{\hat{e}}_s|$.
The transverse wavevector ${\bf k}_{||}$ is constrained by $-\pi/L_x \leq k_x \leq \pi/L_x$ and $-\pi/L_y \leq k_y \leq \pi/L_y$, which define the domain $\mathbb{B}$ known as the first Brillouin zone \cite{Ashcroft76,Kittel96} of the reciprocal lattice ($n=m=0$), the other Brillouin zones are defined by $n, m\not =0$. The symbol
$\rightleftharpoons$ indicates forwards ($\rightarrow$) or backwards ($\leftarrow$) propagation. The corresponding longitudinal wavevector $\pm k^{(c)}_{z,n m}$ is automatically obtained from the value of the frequency $\omega_l=i \xi_{l}$ and transverse wavevector, $\mathbf{ K}_{||}$, together with the appropriate cavity material dispersion relation \cite{Jackson75}
\begin{equation}
-\epsilon_{c}(i \xi_l)\frac{\xi_l^2}{c^2}=\mathbf{K}^{2} _{||,n m} + [k^{(c)}_{z,n m}]^{2} .
\end{equation}
Here ${\rm Re}\, k^{(c)}_{z,nm}+{\rm Im}\, k^{(c)}_{z,nm}>0$ and $\epsilon_{c}(\omega)$ is the permittivity of the medium in the cavity formed by the two gratings.
The two coordinate basis are indeed connected by a transformation which is equivalent to a three dimensional Fourier series and we have
\begin{equation}
\langle \mathbf{r}\vert\mathbf{ k}_{||}, n m,\rightleftharpoons,\lambda \rangle= e^{\imath \mathbf{K}_{||,n m} \cdot \mathbf{R}\pm\imath  k^{(c)}_{z,nm}\, z} ,
\end{equation}
where $\mathbf{R}=(x,y)$. The basis is also orthonormal.

All operators in  \eqref{casimir:energy} can now be represented as (infinite) matrices. Due to the periodicity and the symmetry, each operator only connects wave vectors belonging to different Brillouin zones. Let us consider that the object ``1'' is located to the left of the object ``2'' (i.e. if $z_{1}<z_{2}$ since both gratings are parallel to the $(x,y)$ plane). The reflection operators take the form
\begin{equation}
\mathcal{S}_{1}=
\begin{pmatrix}
\underleftarrow{\mathcal{R}}_{1}& \underrightarrow{\mathcal{T}}_{1}\\
\underleftarrow{\mathcal{T}}_{1} & \underrightarrow{\mathcal{R}}_{1}
\end{pmatrix},
\quad
\mathcal{S}_{2}=
\begin{pmatrix}
\underleftarrow{\mathcal{R}}_{2}& \underrightarrow{\mathcal{T}}_{2}\\
\underleftarrow{\mathcal{T}}_{2} & \underrightarrow{\mathcal{R}}_{2}
\end{pmatrix} .
\end{equation}
Denoting by  $\underleftarrow{\underrightarrow{\mathcal{O}}}$ a generic block matrix, we then have 
\begin{equation}
\langle \mathbf{ k}_{||},   n m, \lambda |\underleftarrow{\underrightarrow{\mathcal{O}}}| \mathbf{ k}'_{||},   n' m',\lambda'\rangle 
=\underleftarrow{\underrightarrow{\mathcal{O}}}^{\lambda;\lambda'}_{ n m; n' m'} \delta(\mathbf{ k}_{||} - \mathbf{ k}'_{||}) .
\end{equation}
Because of the symmetry, in this basis translation operators behave in a very similar way 
\begin{equation}
\mathcal{X}_{12}=
\begin{pmatrix}
0& \underrightarrow{\mathcal{X}}\\
0& 0
\end{pmatrix},
\quad
\mathcal{X}_{21}=
\begin{pmatrix}
0& 0\\
\underleftarrow{\mathcal{X}}& 0
\end{pmatrix} ,
\end{equation}
where $\underleftarrow{\underrightarrow{\mathcal{X}}}$ are diagonal matrices with elements
\begin{equation}
\underleftarrow{\underrightarrow{\mathcal{X}}}^{\lambda}_{ n m}(a)
=- e^{\pm\imath \mathbf{K}_{||,n m} \cdot \mathbf{R}-  \kappa^{(c)}_{z,nm}\, z} ,
\end{equation}
with $\kappa^{(c)}_{z,nm}=-\imath k^{(c)}_{z,nm}= \sqrt{\epsilon_c(i \xi_l) \xi_l^2/c^2 + \mathbf{ K}_{||, nm}^2}$. 

Using that ``${\rm Tr} \log \equiv\log{\rm det}$'' and the Leibniz formula for the determinant of block matrices, Eq.\eqref{casimir:energy} takes the form \cite{Lambrecht06}
\begin{equation}
\mathcal{F}(a) = \frac{1} {\beta} {\sum^{\infty}_{l=0}}'  
\log  {\rm det}\left[1- \underleftarrow{ \mathcal{R}}_{1}\cdot\underrightarrow{\mathcal{X}}(a)\cdot\underrightarrow{\mathcal{R}}_{2} \cdot\underleftarrow{\mathcal{X}}(a)\right] ,
\end{equation} 
or more explicitly 
\begin{widetext}
\begin{equation}
\label{TrLog1BZ}
\mathcal{F}(a) = \frac{L_{x}L_{y}} {\beta} {\sum^{\infty}_{l=0}}' 
\int_\mathbb{B} \frac{{\rm d}^{2} \mathbf{k}_{||}}{(2\pi)^{2}}  
\log  {\rm det}\left[  \delta^{\lambda;\lambda'}_ {n m; n' m'}-\underleftarrow{ \mathcal{R}}_{1}(\mathbf{ k}_{||}, i \xi_l)^{\lambda;\lambda''}_{ n m; n'' m''} \underrightarrow{\mathcal{R}}_{2}(\mathbf{ k}_{||}, i \xi_l)^{\lambda'';\lambda'}_{ n'' m''; n' m'} e^{-a (\kappa^{(c)} _{n' m'}+\kappa^{(c)} _{n'' m''} )}    \right]  ,
\end{equation} 
\end{widetext}
where  the Einstein convention was used. Alternatively, one can also first diagonalize the matrix and then  take the sum of the logarithm of each eigenvalue \cite{Cole68}. Eq. \eqref{TrLog1BZ} clearly reduces the problem of the calculation of the Casimir free energy to the determination of the matrix elements 
$\mathcal{R}_{i}(\mathbf{ k}_{||}, i \xi_l)_{ nm; n'm'}^{\lambda\lambda'}$. We will see in the next section that these quantities correspond to the (Rayleigh) reflection coefficients associated with the scattering of the electromagnetic field from an isolated grating.


\section{\label{sec3} Modal Expansion for 2D Gratings }
\label{Modal}

\begin{figure}
\centering
\includegraphics[width=5.5cm]{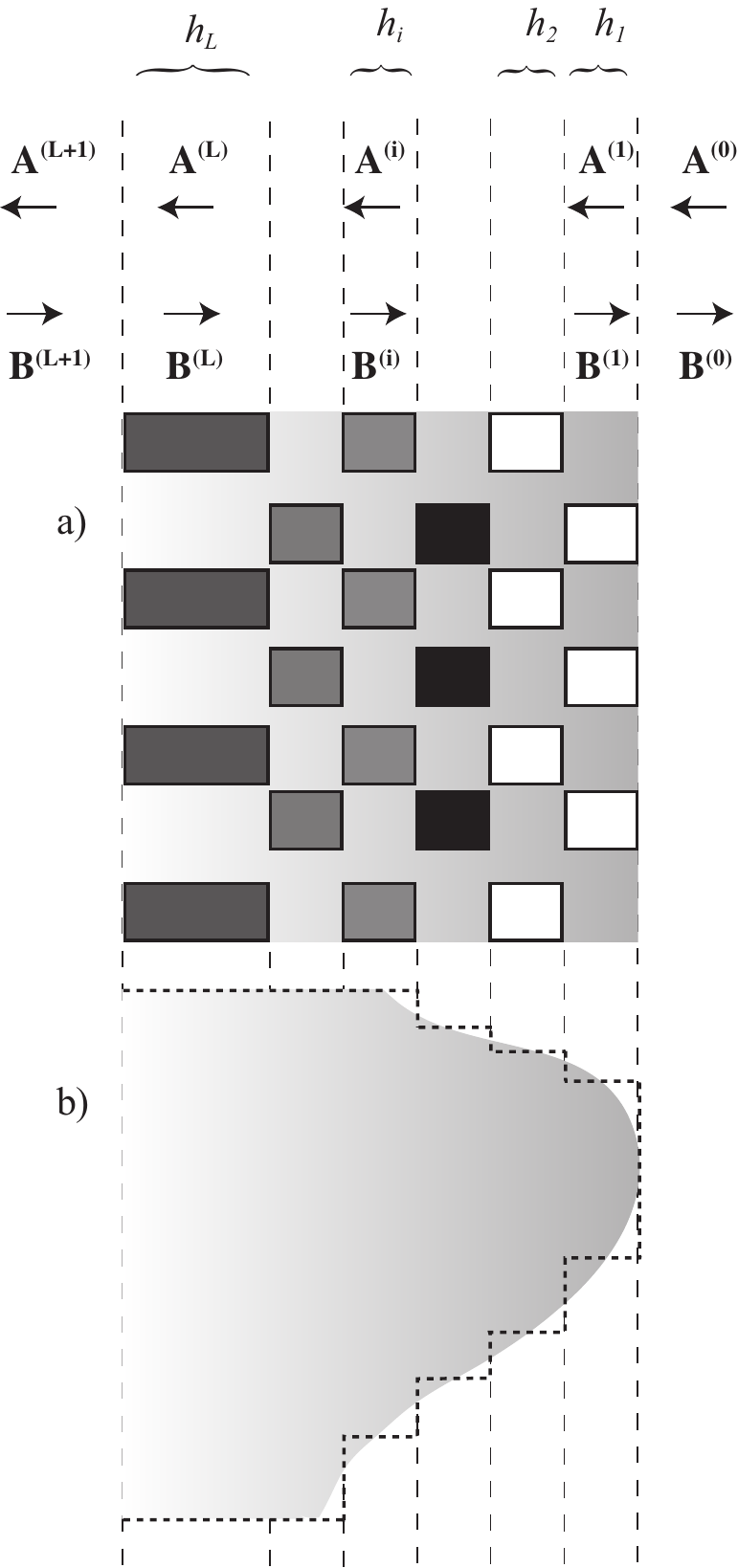}
\caption{ Schematic representation of  a generic periodic  structure.  Each unit cell can be thought of as the result of the superposition of different lamellar layers with the same period. This can be used to calculate scattering properties of arbitrary 3D scatterers provided the period is fixed. An incident planewave in media 0 is scattered into reflected diffration orders, and the transmitted
diffraction orders in $L+1$th media are illustrated.}
\label{fig:nonlamellar}
\end{figure}

Several well developed theoretical and computational techniques exist for the evaluation of scattering from periodic structures \cite{Neviere98}. One class of scattering techniques makes use of the planewave expansion for the electromagnetic fields in the spirit of the derivation of the previous section. These techniques are typically referred to generally as rigorous coupled wave approaches (RCWA) \cite{Johnson11}, or specifically as Fourier modal techniques or planewave techniques. In this section, we will adapt thisl technique to compute the Casimir force from a periodically patterned substrate. 
Each periodic structure will be thought of as the result of the superposition of 2D lamellar gratings (see Fig.\ref{fig:nonlamellar}). For each layer one can derive a direct eigenvalue problem for Maxwell's equations or the associated wave equation \cite{Li97}.  The corresponding complex eigenmodes in the 2D stratum will form a natural basis for the electromagnetic fields in the structure and also a representation of the scattered fields. The scattering properties of the total multilayer medium (lamellar or non lamellar) will be obtained by an iterative application of the scattering matrix following the theory of optical networks. To help the physical intuition
in this section we work with real frequencies $\omega$. However, at the end, the relevant quantities that enter in \eqref{TrLog1BZ} are functions of the imaginary Matsubara frequencies $\omega_{l}=\imath \xi_{l}$. We will discuss the analytical continuation of the modal approach to complex frequencies at the end of this section.


\subsection{The eigenvalue problem in the modulated region}

To begin with let us consider a single layer 2D periodic structure described by the complex dielectric permittivity $\epsilon(x,y)$ and the complex magnetic permeability $\mu(x,y)$, both assumed to be scalars for simplicity. The permittivity and the permeability are periodic functions with period $L_{x}$ and $L_{y}$ of the transverse coordinates $(x,y)$. Our eigenmode calculation starts by decomposing Maxwell's equations into transverse and longitudinal components and follows the approach of L. Li \cite{Noponen94,Li93a,Li97}. This splitting  yields the so called waveguide equations 
\begin{subequations}
\begin{equation}
-\imath k_{0} \partial_{z} \mathbf{E}_{\rm t}
=  \mathbf{\nabla}_{\rm t} 
\left[\chi\,  \mathbf{\hat{z}} \cdot\mathbf{\nabla}_{\rm t}\times \mathbf{H}_{\rm t}\right]  
- k_{0}^2 \mu\, \mathbf{\hat{z}} \times \mathbf{H}_{\rm t} 
\label{eq:1},
\end{equation}
\begin{equation}
-\imath k_{0} \partial_{z} \mathbf{H}_{\rm t}  
 =  -\mathbf{\nabla}_{\rm t} \left[ \zeta\, \mathbf{\hat{z}} \cdot \mathbf{ \nabla}_{\rm t}\times \mathbf{E}_{\rm t} \right]  
 + k_{0}^2 \epsilon\, \mathbf{\hat{z}} \times \mathbf{E}_{\rm t},
\label{eq:2}
\end{equation}
\label{eq:1e2}
\end{subequations}
with $\chi(x,y) =1/\epsilon(x,y)$,  $\zeta(x,y) = 1/\mu(x,y)$, and $k_{0}=\omega/c$. We assume that all fields depend on time as $e^{-i \omega t}$.
The longitudinal electric $E_{z}$ and magnetic  $H_{z}$ fields are not independent and can be determined from the transverse components 
\begin{equation}
E_z = \frac{i}{k_{0}\epsilon} \,  \mathbf{\hat{z}} \cdot\mathbf{ \nabla_t}\times  \mathbf{H}_t
\quad\text{and}\quad H_z = -\frac{i}{k_{0}}\,  \mathbf{\hat{z}} \cdot \mathbf{ \nabla_t}\times  \mathbf{E}_t.
\end{equation}
Eqs. (\ref{eq:1}) and (\ref{eq:2}) further imply that the displacement field and magnetic induction are divergence free.  We have not made any assumptions regarding the transverse mode structure in Eqs.(\ref{eq:1}) and (\ref{eq:2}) since in general the modes will not be classified as,
e.g., transverse electric (TE) or transverse magnetic (TM) in a general complex metallic/dielectric composite structure.

According to the Floquet-Bloch theorem inside each layer the field is a pseudoperiodic function and can be decomposed as follows
\begin{equation}
\mathbf{f}(\mathbf{r})  = \sum_{nm}\mathbf{f}_{nm}(z) \; e^{\imath \mathbf{K}_{||,n m}\cdot \mathbf{R} }
\label{eq:3},
\end{equation}
where we have combined the transverse components of the electric and magnetic fields into a single vector
\begin{equation}
\mathbf{f}(\mathbf{r}) = \left( \begin{array}{r} 
E_{x}(\mathbf{r})    \\
E_{y}(\mathbf{r})  \\
H_{x}(\mathbf{r}) \\
H_{y}(\mathbf{r}) \\
\end{array}   \right),\quad
\mathbf{f}_{nm}(z) = \left( \begin{array}{r} 
E_{x,nm} (z)   \\
E_{y,nm} (z) \\
H_{x,nm} (z)\\
H_{y,nm} (z)\\
\end{array}   \right) ,
\label{eq:8}
\end{equation}
where $E_{x(y),nm}$ and $H_{x(y),nm}$ are the Fourier coefficients of the transverse electric and magnetic fields, respectively. 
We similarly decompose the dielectric permittivity, magnetic permeability and their inverses in Fourier series 
\begin{eqnarray}
\epsilon(x,y)  &=&  \sum_{nm} {\epsilon}_{nm} e^{ i 2\pi n x / L_x + i 2\pi m y/L_y }  , \\
\mu(x,y)  &=&  \sum_{nm} {\mu}_{nm} e^{ i 2\pi n x / L_x + i 2\pi m y/L_y }    , \\
\chi(x,y) &=&  \sum_{nm} {\chi}_{nm} e^{ i 2\pi n x / L_x + i 2\pi m y/L_y }  , \\
\zeta(x,y) &=&  \sum_{nm} {\zeta}_{nm} e^{ i 2\pi n x / L_x + i 2\pi m y/L_y }  .
\end{eqnarray}
Collecting all the Fourier coefficients $\mathbf{f}_{nm}$ in one single large vector $\mathbf{F}$, it is possible to show that the waveguide equations \eqref{eq:1e2} can be recast as a first order matrix differential equation \cite{Noponen94,Cole68} 
\begin{equation}
-ik_{0} \partial_{z} \mathbf{F}(z) =  \mathcal{H}\cdot \mathbf{F}(z) . 
\label{eq:7}
\end{equation}
The matrix $\mathcal{H}$ has a block form and each bloch element is given by \cite{Li97}
\begin{widetext}
{\small
\begin{equation}
\mathcal{H}_{nm:n'm'} =
\begin{pmatrix} 
0 &  0 &  \alpha_{n'}\beta_m \chi^{[nn']}_{[mm']}   & -\alpha_{n} \alpha_{n'} \chi^{[nn']}_{[mm']} + k_{0}^2 \mu^{[nn']}_{[mm']} \\
0 &  0 &  \beta_{m}\beta_{m'}\chi^{[nn']}_{[mm']} - k_{0}^2 \mu^{[nn']}_{[mm']} & -\beta_{m}\alpha_{n'} \chi^{[nn']}_{[mm']} \\
-\alpha_{n} \beta_{m'} \zeta^{[nn']}_{[mm']} &  \alpha_{n} \alpha_{n'} \zeta^{[nn']}_{[mm']}- k_{0}^2 \epsilon^{[nn']}_{[mm']} &  0 & 0 \\
-\beta_{m}\beta_{m'} \zeta^{[nn']}_{[mm']} + k_{0}^2 \epsilon^{[nn']}_{[mm']}  &  \beta_{m} \alpha_{n'} \zeta^{[nn']}_{[mm']}  &  0 & 0 \\
\end{pmatrix}
\label{eq:9} 
\end{equation}
}
\end{widetext}
where $\alpha_{n} = k_x + 2\pi n/L_x$ and $\beta_{m} = k_y + 2\pi m/L_y$. 
The symbol $(\cdot)^{[nn']}_{[mm']}$ means that we have to take the shifted Fourier expansions of the complex permittivity, permeability, and their inverses 
(e.g. $\epsilon^{[nn']}_{[mm']} \equiv \epsilon_{n-n',m-m'}$) that derive from the application of the Laurent rule \cite{Li96a}.
The solution of the first order matrix differential equation \eqref{eq:7} has the form
\begin{equation}
 \mathbf{F} (z)= \mathbf{Y} \exp(\imath k_{0} \gamma z)
\label{eq:10},
\end{equation}
where $\mathbf{Y}$ and $\gamma$ are respectively one of the eigenvectors and the corresponding eigenvalue that are solutions to the eigevalue problem \cite{Cole68,Naimark68}
\begin{equation}
\gamma k_{0}^2   \mathbf{Y} = \mathcal{H}\cdot  \mathbf{Y} .
\label{eq:eigenmode}
\end{equation}
Note that the value of the transverse momentum in the first Brillouin zone $\mathbf{k}_{||}$ is fixed in these equations, so that the set of eigenvectors $\mathbf{Y}_{\nu}$ and corresponding eigenvalues $\gamma_{\nu}$ are functions of $\mathbf{k}_{||}$ .
Given the fact that $\mathcal{H}$ is non-hermitian, the eigenvalues are in general complex and the eigenvectors are not orthogonal \cite{Cole68,Naimark68} to each other. However the eigenvectors are \emph{bi-orthogonal}  \cite{Cole68,Naimark68} to the eigenvectors of the corresponding adjoint equation
\begin{equation}
\lambda k_{0}^2   \mathbf{Y}^{\dag} =\mathbf{Y}^{\dag}\cdot\mathcal{H}^{\dag} .
\label{eq:left}
\end{equation}
From the theory of non-self adjoint differential equations one knows that the eigenvalues and eigenvectors of the two mutually adjoint equations can be ordered in such a way that 
$\lambda_{\nu}=\gamma^{*}_{\nu}$. Defining the scalar product
\begin{equation}
\langle {\boldsymbol \psi}| {\boldsymbol \varphi} \rangle\equiv \sum_{nm} \psi^{*}_{nm}\varphi_{nm} ,
\end{equation}
we have that
\begin{equation}
\langle  \mathbf{Y}^{\dag}_{\nu} |  \mathbf{Y}_{\nu'}\rangle =    \delta_{\nu,\nu'} .
\end{equation}
Note that the dimension of the matrix $\mathcal{H}$ is even 
\begin{equation}
{\rm dim}[\mathcal{H}] = 2D\times 2D,
\end{equation}
where $D=2(2N+1)(2M+1)$ and $2N+1$, $2M+1$ are the number of Fourier terms in the series expansion along the $x$ and $y$ directions, respectively. 
The characteristic equation for the eigenvalues is an equation of an even order with $2D$ solutions: if $\gamma_{\nu}$ is a solution then $\gamma_{\nu}^{*}$ is also
a solution, so that half of the eigenvalues and half of the eigenvectors can be obtained by simple conjugation. Notation-wise it is also possible to associate to the same value 
$\gamma_{\nu}$ one eigenvector of eq.\eqref{eq:7} and one eigenvector of the adjoint equation, i.e.
\begin{equation}
\gamma_{\nu}\leftrightarrow 
 \mathbf{Y}_{\nu},\mathbf{Y}^{\dag *}_{\nu} ,
\end{equation} 
and, in literature, $\mathbf{Y}_{\nu}$ and  $\mathbf{Y}^{\dag *}_{\nu}$ are generally called right-handed and left-handed eigenvector respectively (the \emph{bi-orthogonality} property can also be reinterpreted in terms of the previous definition). 
Furthermore, if we rearrange the vector ${\bf F}(z)$ to have the form
\begin{equation}
\mathbf{F}(z)=
\begin{pmatrix}
\mathbf{F}_{\mathbf{E}}(z)\\
\mathbf{F}_{\mathbf{H}}(z)
\end{pmatrix} ,
\end{equation}
where ${\bf F_E}$  and ${\bf F_H }$ contains all electric field and magnetic field components, respectively.Then the matrix $\mathcal{H}$ takes the block form
\begin{equation}
\mathcal{H}=
\begin{pmatrix}
0& \mathcal{H}_{\mathbf{H}}\\
\mathcal{H}_{\mathbf{E}} & 0
\end{pmatrix} .
\end{equation}
The eigenvalue problem \eqref{eq:eigenmode} can be reduced to separate equations \cite{Li97}
\begin{subequations}
\begin{gather}
\gamma^{2}k_{0}^{2}\mathbf{Y}_{\mathbf{E}}= \mathcal{H}_{\mathbf{H}}\cdot\mathcal{H}_{\mathbf{E}}\cdot\mathbf{Y}_{\mathbf{E}} , \\
\gamma^{2}k_{0}^{2}\mathbf{Y}_{\mathbf{H}}= \mathcal{H}_{\mathbf{E}}\cdot\mathcal{H}_{\mathbf{H}}\cdot\mathbf{Y}_{\mathbf{H}} ,
\end{gather}
\end{subequations}
so that we can deduce that the eigenvalues come in pairs and so do the eigenvectors
\begin{equation}
\gamma_{\nu} \leftrightarrow \underrightarrow{\mathbf{Y}}_{\nu}\text{ and } 
-\gamma_{\nu}  \leftrightarrow \underleftarrow{\mathbf{Y}}_{\nu} ,
\end{equation}
representing forward ($\rightarrow$) and backward ($\leftarrow$) propagation, respectively.

These considerations ensure that the previous eigenmodes are \emph{natural basis} for the electromagnetic field in the modulated region. Collecting all these
results and definitions we have
\begin{equation}
\mathbf{F}(z) =\sum_{\nu}  \underleftarrow{\mathbf{Y}}_{\nu}(\mathbf{k}_{||}) e^{-\imath k_{0} \gamma_{\nu} z}   A_{\nu} +
 \underrightarrow{\mathbf{Y}}_{\nu}(\mathbf{k}_{||}) e^{ \imath k_{0} \gamma_{\nu} z} B_{\nu} 
\label{eq:expansion1}.
\end{equation}
Here we have explicitly indicated the dependency of the eigenvectors on $\mathbf{k}_{||}$. The expression for the total field $\mathbf{f}(\mathbf{r})$ can be obtained by isolating the four dimensional vectors $\mathbf{f}_{nm}$ from the previous expression and inserting them in Eq.\eqref{eq:3}.
The only unknowns in Eq.\eqref{eq:expansion1} are the coefficients $A_{\nu}$ and $B_{\nu}$, the field amplitudes (scalars), and they will be completely determined in the next subsection once we impose the boundary conditions on the fields at the border of each modulated region.
Note that the Rayleigh basis defined in Section \ref{Casimir} for the cavity is a particular (homogeneous medium / no modulation) case of the modal solution described above, and the 
eigenvalues coincide with the longitudinal vector $k^{(c)}_{z,n m}$, with multiplicity two (one for each polarization $\lambda=s, p$). This correspondence means that we can safely use the same formalism throughout the structure.


\begin{figure}
\centering\includegraphics[width=5.5cm]{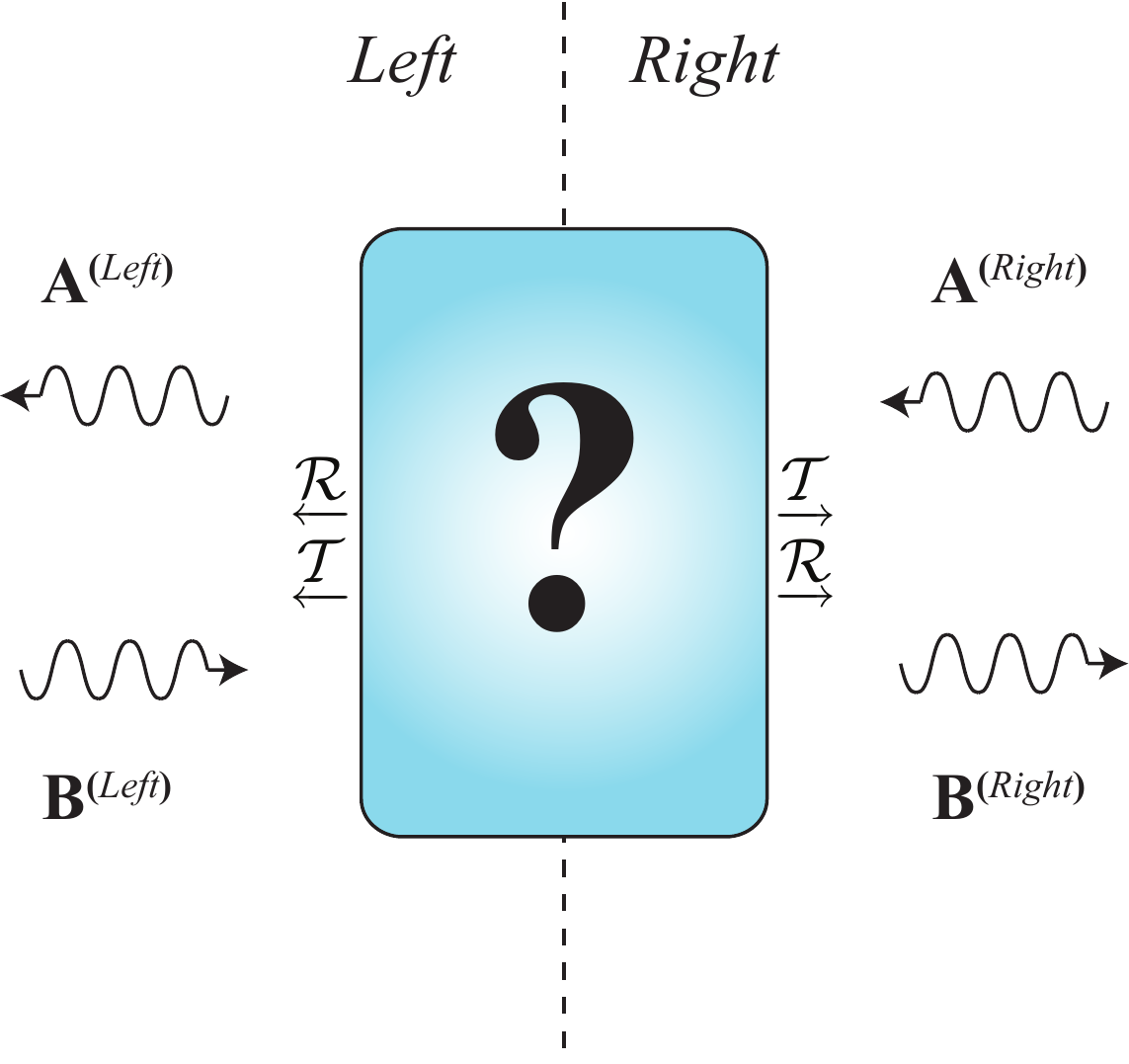}
\caption{Schematic description of the scattering process by a generic object. The scattering matrix formalism connects the output field amplitudes as a function of the input field amplitudes for a general scattering object.}
\label{fig:network}
\end{figure}

\subsection{Scattering and boundary conditions}

Let us now consider the problem of the scattering of an electromagnetic wave by a periodic structure in its completeness, and assume that the network element of Fig.\ref{fig:network} is composed by several 2D periodic layers with thickness $h_{i}$, electric permittivity
$\epsilon_i(x,y)$, and magnetic permeability $\mu_i(x,y)$
($i=1\dots L$, see Fig.\ref{fig:nonlamellar}). The first ($1|0$) interface is at the right of the network element at position $z=0$, while the last ($L+1|L $) is at the left of the network at $z=-h=\sum_{i}h_{i}$. The first ($\epsilon_{0},\mu_{0}$ and $z>0$) and the last media ($\epsilon_{L+1},\mu_{L+1}$ and $z<-h$) are assumed to be uniform layers with planewave type eigenmodes (Rayleigh basis). For each layer we can write the expression given in Eq.\eqref{eq:expansion1} in the following form
\begin{equation}
\mathbf{F}^{(i)}(z) =\mathcal{Y}^{(i)}\cdot t^{(i)}_{\rm pr} \cdot \left( \begin{array} {c} \mathbf{A}^{(i)} \\ \mathbf{B}^{(i)} \end{array} \right) 
\label{eq:expansionlayer}~.
\end{equation}
The matrix $\mathcal{Y}^{(i)}$ contains all eigenvectors as columns
\begin{equation}
\mathcal{Y}^{(i)}\equiv\left( \underleftarrow{\mathbf{Y}}^{(i)}_{\nu=1},\dots,\underleftarrow{\mathbf{Y}}^{(i)}_{\nu=D},\underrightarrow{\mathbf{Y}}^{(i)}_{\nu=1},\dots,\underrightarrow{\mathbf{Y}}^{(i)}_{\nu=D}\right).
\end{equation}
The block matrix
\begin{equation}
t^{(i)}_{\rm pr}=
\begin{pmatrix}
\boxed{e^{\imath k_{0}\gamma^{(i)}  h_i} } & 0\\
0& \boxed{e^{-\imath k_{0}\gamma^{(i)}  h_i}}
\end{pmatrix}
\label{eq:transfer}
\end{equation}
has the exponential functions with the eigenvalues $\gamma^{(i)}_{\nu}$ along the diagonal of each of the two $D\times D$ block submatrices and describes the propagation through the layer of thickness $h_{i}$. The field amplitudes for each layer $A^{(i)}_{\nu}$, $B^{(i)}_{\nu}$ are collected in the column vectors $\mathbf{A}^{(i)}$ and $\mathbf{B}^{(i)}$. 
That means that the vectors $\mathbf{A}^{(0)}$  and  $\mathbf{B}^{(0)}$ with components $A^{(0)}_{\nu}$ and $B^{(0)}_{\nu}$ give the incident and reflected field amplitudes from the right of the network. Similarly $\mathbf{A}^{(L+1)}$ and $\mathbf{B}^{(L+1)}$ are the reflected and incident fields from the left of the network.
The right and left reflection matrices are then defined as (see Fig. \ref{fig:network})
\begin{equation}
\mathbf{B}^{(0)}=\underleftarrow{\mathcal{R}}\cdot \mathbf{A}^{(0)}, \quad \mathbf{A}^{(L+1)}=\underrightarrow{\mathcal{R}}\cdot \mathbf{B}^{(L+1)} .
\end{equation}
Similarly the left and right transmission operators can be defined  as follows
\begin{equation}
\mathbf{A}^{(L+1)}=\underleftarrow{\mathcal{T}}\cdot \mathbf{A}^{(0)}, \quad \mathbf{B}^{(0)}=\underrightarrow{\mathcal{T}}\cdot \mathbf{B}^{(L+1)}.
\end{equation}

Imposing the continuity of the tangential fields at the $z=z_{i}=-\sum_{j=1}^{i}h_{j}$ interface means that $
\mathbf{F}^{(i+1)}(z_{i})=\mathbf{F}^{(i)}(z_{i})$ and the transfer operator that relates the field amplitudes in the $i$th and  $i+1$th layer is defined as 
\begin{align}
\left( \begin{array} {c} \mathbf{A}^{(i+1)} \\ \mathbf{B}^{(i+1)} \end{array} \right)
=t^{(i+1|i)} \cdot \left( \begin{array} {c} \mathbf{A}^{(i)} \\ \mathbf{B}^{(i)} \end{array} \right) ,
\end{align}
where 
\begin{equation}
t^{(i+1|i)}=
\begin{pmatrix}
t^{(i+1|i)}_{11}&t^{(i+1|i)}_{12}\\
t^{(i+1|i)}_{21}&t^{(i+1|i)}_{22}\\
\end{pmatrix} .
\end{equation}
Each single block of the transfer matrix has a dimension $D\times D$. Defining rectangular $D\times2D$ matrices
\begin{equation}
\underleftarrow{\underrightarrow{\mathcal{Y}}}^{(i)}\equiv(\underrightarrow{\underleftarrow{\mathbf{Y}}}^{(i)}_{\nu=1},\dots,  \underrightarrow{\underleftarrow{\mathbf{Y}}}^{(i)}_{\nu=D})
\end{equation}
the block elements can be expressed as overlaps of the mode eigenvectors
\begin{eqnarray}
t^{(i+1|i)}_{11}  & =  & \underleftarrow{\mathcal{Y}}^{(i+1)\dag}\cdot\underleftarrow{\mathcal{Y}}^{(i)}  ,
\nonumber \\ 
t^{(i+1|i)}_{12}  & =  &  \underleftarrow{\mathcal{Y}}^{(i+1)\dag}\cdot\underrightarrow{\mathcal{Y}}^{(i)} ,
\nonumber \\ 
t^{(i+1|i)}_{21}  & =  &  \underrightarrow{\mathcal{Y}}^{(i+1)\dag}\cdot\underleftarrow{\mathcal{Y}}^{(i)} ,
\nonumber \\ 
t^{(i+1|i)}_{22}  & =  & \underrightarrow{\mathcal{Y}}^{(i+1)\dag}\cdot\underrightarrow{\mathcal{Y}}^{(i)} .
\end{eqnarray}
The field at the point $z=z_{i+1}$ is related to that at $z=z_{i}$ by 
\begin{equation}
\mathbf{F}^{(i+1)}(z_{i+1})=\theta^{(i+1|i)}\mathbf{F}^{(i)}(z_{i}) ,
\end{equation}
with 
\begin{equation}
\theta^{(i+1|i)}=t^{(i+1|i)}\cdot t^{(i)}_{\rm pr} 
\end{equation}
being the total transfer matrix from layer $i$ to layer $i+1$.

Now we could construct the transfer matrix of the whole structure by iterating through the multilayer (lamellar or non-lamellar) structure, and then solve for the reflection operator.
However, if numerically implemented, the transfer matrix method is known to suffer from instabilities when the layers are thick, due to the growing exponentials  contained in the transfer matrix \cite{Li96}. A remedy to these numerical 
instabilities is the S-matrix approach. The S-matrix is derived from the ordered T-matrix, where the ordering is 
in terms of forward and back propagating modes (see Fig.\ref{fig:network}).   
The  layer S-matrix can be defined by reordering the coefficients, and can be expressed 
in terms of the interface transfer matrices
\begin{widetext}
\begin{equation}
\left( \begin{array} {c} \mathbf{A}^{(i+1)} \\ \mathbf{B}^{(i)} \end{array} \right) = 
\left( \begin{array}{cc} 
1 &  0  \\
0 &  e^{\imath k_{0}\gamma^{(i)}  h_i}  \\
\end{array}   \right)\cdot 
\left( \begin{array}{cc} 
\sigma^{(i+1|i)}_{11} &  \sigma^{(i+1|i)}_{12}   \\
\sigma^{(i+1|i)}_{21} &  \sigma^{(i+1|i)}_{22}   \\
\end{array}   \right)\cdot
\left( \begin{array}{cc} 
e^{\imath k_{0}\gamma^{(i)}  h_i} &  0  \\
0 &  1  \\
\end{array}   \right) 
\cdot \left( \begin{array} {c} \mathbf{A}^{(i)} \\ \mathbf{B}^{(i+1)} 
\end{array} \right)\equiv\Sigma^{(i+1|i)}\cdot \left( \begin{array} {c} \mathbf{A}^{(i)} \\ \mathbf{B}^{(i+1)} \end{array} \right) .
\end{equation}
The interface S-matrix is 
\begin{equation}
\left( \begin{array}{cc} 
\sigma^{(i+1|i)}_{11} &  \sigma^{(i+1|i)}_{12}   \\
\sigma^{(i+1|i)}_{21} &  \sigma^{(i+1|i)}_{22}   \\
\end{array}   \right)  = 
\left( \begin{array}{cc} 
t^{(i+1|i)}_{11} - t^{(i+1|i)}_{12}  [ t^{(i+1|i)}_{22}]^{-1} t^{(i+1|i)}_{21} &  t^{(i+1|i)}_{12}   [t^{(i+1|i)}_{22}]^{-1} \\
- [ t^{(i+1|i)}_{22}]^{-1}  t^{(i+1|i)}_{21} &  [t^{(i+1|i)}_{22}]^{-1}  \\
\end{array}   \right) .
\label{smat-1}
\end{equation}
\end{widetext}
The S-matrix stability is guaranteed in Eq. (\ref{smat-1}) due to the exponential decay (${\rm Im}\gamma^{i}_{\nu}>0 $) of the backward propagating mode. The layer S-matrix
\begin{equation}
\Sigma^{(i+1|i)}= 
\left( \begin{array}{cc} 
s^{(i+1|i)}_{11} &  s^{(i+1|i)}_{12}   \\
s^{(i+1|i)}_{21} &  s^{(i+1|i)}_{22}   \\
\end{array}   \right) 
\end{equation}
incorporates the propagation phase factor and the interface S-matrices. Again, the S-matrix of the multilayer object can be constructed using an iterative procedure with the following recursion relations 
\begin{widetext}
\begin{eqnarray}
\Sigma^{(i+1|0)}_{11} & = & s^{(i+1|i)}_{11}\left( 1-s^{(i+1|i)}_{21}\Sigma^{(i|0)}_{12} \right)^{-1}\Sigma^{(i|0)}_{11} , \nonumber \\ 
\Sigma^{(i+1|0)}_{12} & = & s^{(i+1|i)}_{12} +s^{(i+1|i)}_{11}\left( 1-s^{(i+1|i)}_{21}\Sigma^{(i|0)}_{12} \right)^{-1} s^{(i+1|i)}_{22} \Sigma^{(i|0)}_{12} , \nonumber \\ 
\Sigma^{(i+1|0)}_{21} & = & \Sigma^{(i|0)}_{21} + \Sigma^{(i|0)}_{22}\left( 1-s^{(i+1|i)}_{21}\Sigma^{(i|0)}_{12}\right)^{-1} s^{(i+1|i)}_{21}\Sigma^{(i|0)}_{11} , \nonumber \\ 
\Sigma^{(i+1|0)}_{22} & = & s^{(i+1|i)}_{22} \Sigma^{(i|0)}_{22}\left( 1-s^{(i+1|i)}_{21}\Sigma^{(i|0)}_{12}\right)^{-1}  .
\end{eqnarray}
\end{widetext}
It is clear that the scattering matrix of the multilayer scatterer which connects the incident media (incident and reflected modes) to the exit media (transmitted modes) is given by  $\Sigma^{(L+1|0)}$.
%
Unfortunately the definition of the scattering matrix in the grating literature \cite{Li97} does not match the one used in the network theory (see Section \ref{Casimir}). However this last one can be obtained by a simple reordering of the block matrices
\begin{equation}
\mathcal{S}=
\begin{pmatrix}
\Sigma_{21}&  \Sigma_{22}\\
\Sigma_{11}& \Sigma_{12}
\end{pmatrix} .
\end{equation}
From here all the relevant reflection (and transmission) operators for the calculation of the Casimir free energy can be obtained.




\subsection{Analytical properties of the modal approach}

In the previous section we derived the scattering matrix for a general 2D-periodic structure having in mind a real value for the frequency of the field.  For the calculation of the Casimir free energy it is more  convenient to work with imaginary frequencies from the very beginning, since the quantities entering in Eq.\eqref{TrLog1BZ} are functions of $\omega=\imath \xi$. From an analysis of the previous section one can easily realize that the consequence of a (Wick) rotation in the complex plane (from the real frequency axis to the imaginary frequency axis) is automatically connected with the analytical properties of the involved functions. Maxwell waveguide equations \eqref{eq:1e2} can be easily written in terms of imaginary frequency. The dielectric permittivity and the magnetic permeability are analytic functions in upper part of the complex plane with the property \cite{Jackson75}
\begin{equation}
\epsilon(z)=\epsilon^{*}(-z^{*}),\quad \mu(z)=\mu^{*}(-z^{*}),
\end{equation}
from which it immediately follows that $\epsilon(\imath\xi)$ and $\mu(\imath\xi)$ are real. Since we are dealing with passive media they also are positive quantities for $\xi>0$. Clearly now the matrix $\mathcal{H}$ is real. By looking at the elements of $\mathcal{H}$ one can easily show  that its characteristic equation $\Delta_{2D}(z,\alpha,\beta,\gamma^{2})=0$ inherits some of the properties of the permittivity and of the permeability, in particular it is true that
\begin{equation}
\Delta_{2D}(z,\alpha,\beta,\gamma^{2})=\Delta_{2D}(-z^{*},\alpha,\beta,[\gamma^{2}]^{*})
\end{equation}
which means that the eigenvalues satisfy the property
$\gamma^{2}_{\nu}(z,\alpha,\beta)=[\gamma_{\nu}^{2}(-z^{*},\alpha,\beta)]^{*}$.
The choice of the definition of the square root and its continuity in the complex plane finally implies that 
\begin{equation}
\gamma_{\nu}(z,\alpha,\beta)=-\gamma_{\nu}^{*}(-z^{*},\alpha,\beta) ,
\end{equation}
from which it is possible to conclude that $\gamma_{\nu}(\omega,\alpha,\beta)=-\gamma_{\nu}^{*}(-\omega,\alpha,\beta)$, and that $\gamma_{\nu}(\imath\xi,\alpha,\beta)$ ($\xi \in {\rm Re}$) is pure imaginary. Since they depend only upon the sign of the eigevalues,  the concepts of forward and backward propagation can be easily generalized and we can still write an expansion like \eqref{eq:expansion1}, working with imaginary frequency. The main difference is now that the exponentials are real functions. The formalism of the scattering matrix can be reapplied  to derive the reflection operators at imaginary frequencies. Naturally, in the numeric calculation the necessary truncation of the matrix $\mathcal{H}$ leads to some small deviations from the previous conclusions. For example the eigenvalues as a function of the imaginary frequency can acquire a small real part.

In summary, to calculate the Casimir free energy between periodic structures it is sufficient to write the waveguide equation for each Matsubara frequency and perform the corresponding modal expansion procedure.  The resulting reflection matrices are then inserted in Eq.\eqref{TrLog1BZ} and one can choose between the calculation of the determinant of the final expression, or its diagonalization followed by the calculation of the trace of the resulting diagonal matrix. 
Special attention is required when computing  the zero Matsubara frequency contribution to the free energy $\xi=0$ in the scattering formalism. This will be outlined in the next section.


\subsection{Zero frequency modal solutions}

The computation of the Casimir free energy requires explicitly the $l=0$ (zero Matsubara frequency $\xi_0=0$) scattering matrices from the modal solution to Maxwell's equations.  In the following we will consider non-magnetic media for simplicity, $\mu(x,y)=1$. 
The zero frequency limit is intrinsically tied to dispersive models of the permittivity since in general the product
$\omega \epsilon(\omega; x,y)$ appears explicitly in Maxwell's equations. Therefore, if the permittivity is finite at zero frequency then the electric and magnetic field are irrotational. However, if the permittivity has a simple pole at zero frequency the fields are coupled. 

The simplest and most useful model for the permittivity of a homogeneous dielectric  medium is the oscillator model or Drude-Lorentz model \cite{Jackson75}. This model assumes harmonically bound charges to an ion core that makes up a neutralizing background.  The resulting dielectric permittivity is 
\begin{equation}
\epsilon(\omega) = 1 -  \frac{\Omega^2_{pl}}{\omega^2-\omega_0^2+i\Gamma \omega } ,
\label{osc:eq1}
\end{equation} 
where $\Gamma$ is the damping coefficient, $\omega_0$ is the oscillator frequency, and the plasma frequency is defined as
\begin{equation}
\Omega^2_{pl} = \frac{4\pi e^2 n}{ m}.
\label{plasmafreq}
\end{equation}
This simple model of a dispersive media can be generalized to multiple oscillators with different resonance frequencies, oscillator strengths and linewidths, and forms the basis for models of optical dispersion in dielectrics and metals. The Drude model for a metal is found by considering the limit of free electrons (i.e $\omega_0 \rightarrow 0$ of Eq. (\ref{osc:eq1}) )
and is given by
\begin{equation}
\epsilon(\omega) = 1 -  \frac{\Omega^2_{pl}}{\omega^2+i\Gamma \omega } .
\label{drude:1}
\end{equation} 
The Drude model has two poles, one pole at zero frequency and the other pole in the lower half of the complex plane \cite{Intravaia09}.  
The connection with constituent relationships in Maxwell's equations is obtained by expanding Eq.(\ref{drude:1}),
\begin{equation}
\epsilon(\omega) = 1 -  \frac{i\Omega^2_{pl}}{\Gamma(\omega+i\Gamma)} +  \frac{i\Omega^2_{pl}}{\Gamma\omega} ,
\label{drude:2}
\end{equation} 
and we can identify the first two terms as the Debye permittivity, and the last term with the Drude conductivity, $\sigma = \Omega_{pl}^2/\Gamma$.  In the limit $\Gamma \rightarrow 0$, the Drude conductivity is infinite and we recover the London's superconductor \cite{London35} or also plasma model for the metallic media, which is singular at zero frequency.  It is therefore important the order in which the limits are taken, and this can yield differing results \cite{Milton04,Klimchitskaya09c}.   
In all our calculations, we consider the zero frequency limit first with finite conductivity at zero frequency.

The zero frequency limit to Maxwell's equations (Eq. (\ref{eq:1}) and Eq.(\ref{eq:2})) is 
\begin{equation}
\partial_{z} \mathbf{E}_{\rm t}   =  \mathbf{ \nabla}_{\rm t} \left[\frac{c}{4\pi \sigma} \mathbf{\hat{z}}\cdot\mathbf{\nabla}_{\rm t}\times \mathbf{H}_{\rm t} \right]  
\label{zf:1},
\end{equation}
and
\begin{equation}
 \partial_{z} \mathbf{H}_{\rm t}  =  \mathbf{\nabla}_{\rm t} H_z  - \frac{4\pi \sigma}{c} \mathbf{\hat{z}} \times \mathbf{E}_{\rm t}  ,
\label{zf:2}
\end{equation}
where we have used the Drude model of Eq.(\ref{drude:2}), and $\sigma=\sigma(x,y)$ is the spatially varying DC conductivity of the
2D periodic structure. These two equations replace the waveguide equations  and are the basis for the zero frequency modal solutions. The corresponding transverse fields are the zero frequency limits of the transverse propagating solutions, and as such are different from purely static solutions.  The zero frequency waveguide equations depend on the $z$ component of the magnetic field.  We therefore  consider only transverse magnetic (TM) zero frequency solutions, $H_z=0$, since a non-zero value would imply a DC surface current is flowing, which is inconsistent with a non zero dissipation.  Furthermore, for a planar metallic mirror the zero frequency Fresnel reflection matrix for the transverse electric (TE) mode vanishes, and we are therefore restricted to TM solutions ($H_z=0$) that contribute to the zero Matsubara frequency in the expression for Casimir free energy. As a technical remark we point out that in the regions within the structure containing vacuum the conductivity $\sigma(x,y)$ is obviously
zero, and the equations above are singular. In the numerics, we choose a vanishingly small but non-zero conductivity for those regions. The final results turn out to be independent of this choice. Alternatively, we can consider small but finite frequencies approaching the zero frequency limit using the assumed material dispersion.  Both approaches lead to the same converged zero frequency contribution for the grating structures examined. 


\section{Results}\label{Results}

\begin{figure}
\centering\includegraphics[width=6.5cm]{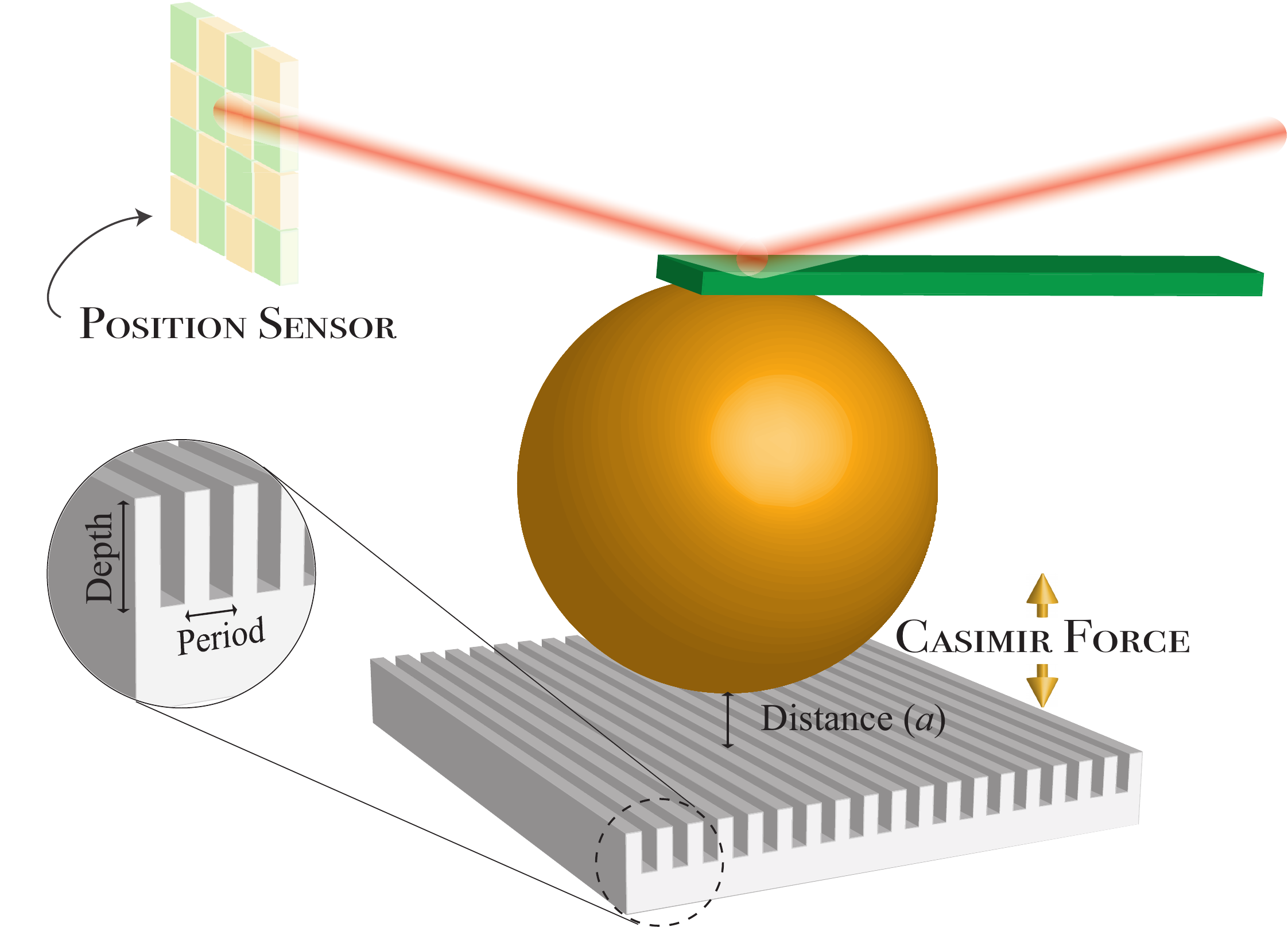}
\caption{AFM setup for measuring Casimir interactions between a sphere and a 1D lamellar grating.}
\label{fig:1}
\end{figure}

Most experiments measuring the Casimir force between two objects use a sphere or a spherical lens as one of the intervening objects. This avoids parallelism issues which could affect the precision required by the measurements (see Fig. \ref{fig:1}).  Although the scattering approach allows one to take into account spherical or even more complicate geometries \cite{Rahi09,Canaguier-Durand09,Canaguier-Durand10,Canaguier-Durand10a,Zandi10}, the modal method presented in this paper is not suited for non-periodic, spherical surfaces.
However,  the radius of curvature $R$ of the sphere (or spherical lens) used in experiments is so much larger than the distance to the other plate that one can safely use
the so-called proximity force approximation (PFA) \cite{Derjaguin56, Bordag10}. In our case PFA relates the Casimir force between a spherical surface and a grating to the free energy per unit of area between a plane and a grating
\begin{equation}
F_{sph-g} (a)= 2\pi R ~ {\mathcal{F}_{pl-g}} ,
\end{equation}
and the force gradient in the sphere-grating configuration is then related to the force per unit area in the plane-grating geometry.
The quantity on the r.h.s. can be calculated starting from the results presented in the previous sections.
In this section we benchmark our finite-temperature numerical code against a recent precise measurement of the Casimir force gradient between a metallic sphere and a deeply etched ($\approx 1$ micron) 1D lamellar silicon grating \cite{Chan08}.  We then move on to more complex 2D periodic structures.

The 1D lamellar grating in the experiment of Ref.\cite{Chan08} was p-doped Si and the sphere was metallized with Au.
In order to have a precise comparison between our numerics and the
experimental data one should input into our code the actual optical properties of the samples used in the experiment. Unfortunately these were not measured in \cite{Chan08}, so
here we will use tabulated optical data for Au and p-doped Si, that have been compiled and studied by several authors  \cite{Arnold79,Bruhl02,Pirozhenko08}.
We model the intrinsic Si permittivity by a Drude-Lorentz model,
\begin{equation}
\epsilon_{Si} (i \xi) = \epsilon_{ \infty}+ (\epsilon_{0}-\epsilon_{ \infty})\frac{\omega_0^2}{\xi^2+\omega_0^2} ,
\end{equation}
with $\epsilon_0 = 11.87$, $\epsilon_{ \infty} = 1.035$ , and $\omega_0 =6.6\times10^{15}$ rad/s. The p-doped Si is modeled by adding
to the intrinsic part a Drude background
\begin{equation}
\epsilon_{doped}(i\xi)  = \epsilon_{Si} (i\xi) + \frac{\omega_p^2}{\xi(\xi +\gamma)},
\end{equation}
with $\omega_p = 3.6151\times10^{14}$ rad/sec and $\gamma = 7.868\times10^{13}$ rad/sec.
Similarly, the Au sphere is modeled by a Drude model
\begin{equation}
\epsilon_{Au} (i\xi) = 1+ \frac{\Omega_p^2}{\xi(\xi +\Gamma)} ,
\end{equation}
with $\Omega_p = 1.27524\times10^{16 }$ rad/sec and $\Gamma = 6.59631\times10^{13 }$ rad/sec.

With the materials parameters at hand, we can begin comparing the experimental data and the computed exact scattering solution for the grating.  Our code
is designed for 2D periodic systems, so in order to treat 1D gratings we use a period along the translational invariant direction (say $y$)
equal to that along the non-invariant direction $x$, i.e. $L_x=L_y$, so that the resulting
structure is 2D periodic. In all our calculations we will truncate all the Fourier series to $N=M=5$.
This implies that the dimension of the reflection matrices is $2(2N+1)^2 \times 2(2N+1)^2 = 242 \times 242$.
The $\mathbf{k}_{||}$-space integration is performed using a 16pt Gauss-Legendre quadrature.
We set the temperature to $T=300$K, and use the first 36 Matsubara frequencies for all the studied range of distances between
the sphere and the grating.

\begin{figure}[tbp]
\includegraphics[width=9cm]{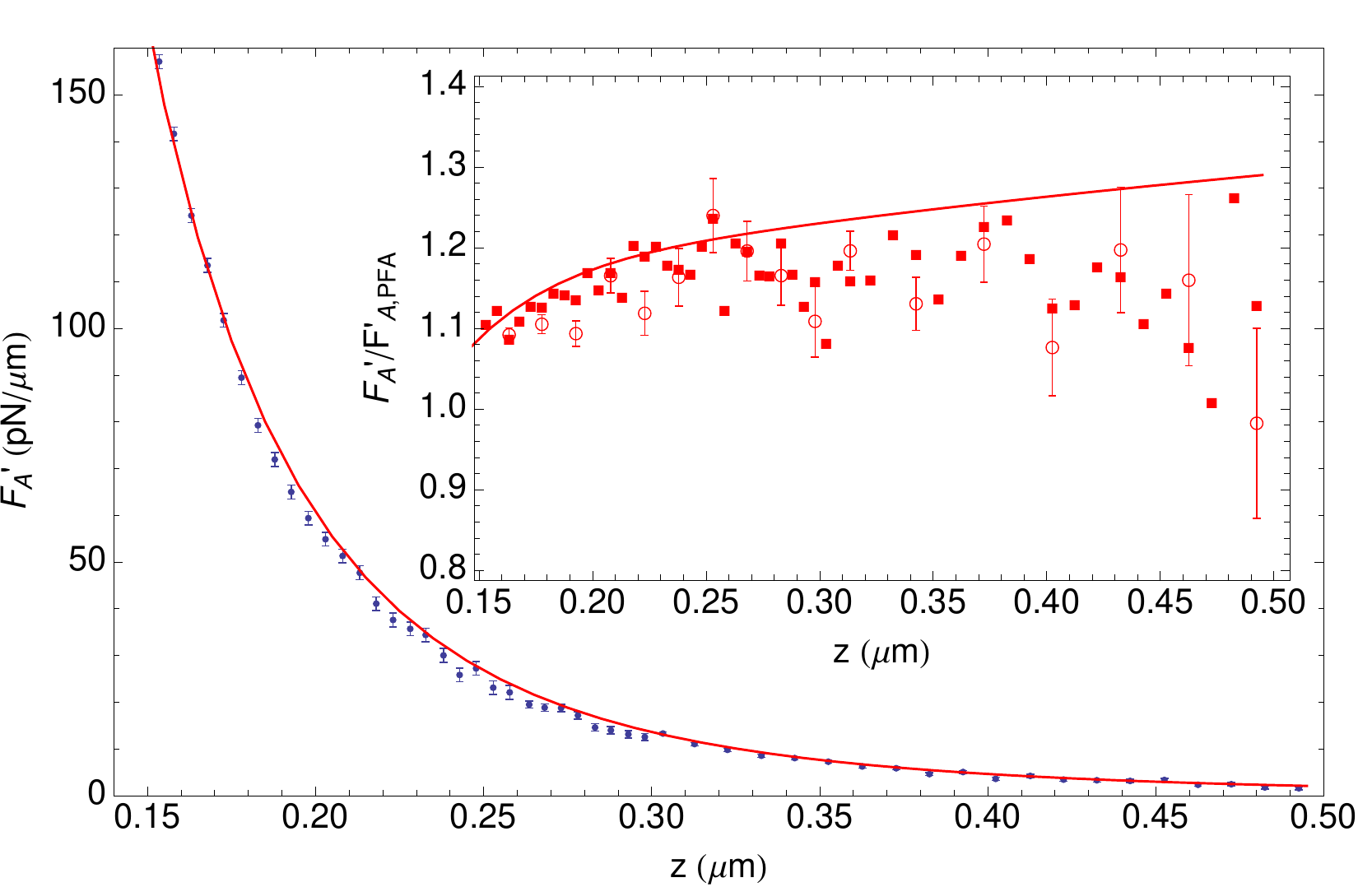}
\caption{Casimir force gradient in the sphere-grating geometry for sample A in \cite{Chan08}: exact numerics (solid) and experimental data  (dots with error bars). The inset shows the ratio of the Casimir force gradient divided by the PFA prediction:  exact numerics/theoretical PFA (solid), exact numerics/``experimental" PFA (squares), and experimental data/``experimental'' PFA (hollow circles with error bars).
The geometrical parameters of grating A are: period=1000 nm, depth=980 nm,
and filling factor=0.510. The radius of the sphere is $50 \mu$m, and the temperature is set to $T=300$K.}
\label{fig:ForceGradientA}
\includegraphics[width=9cm]{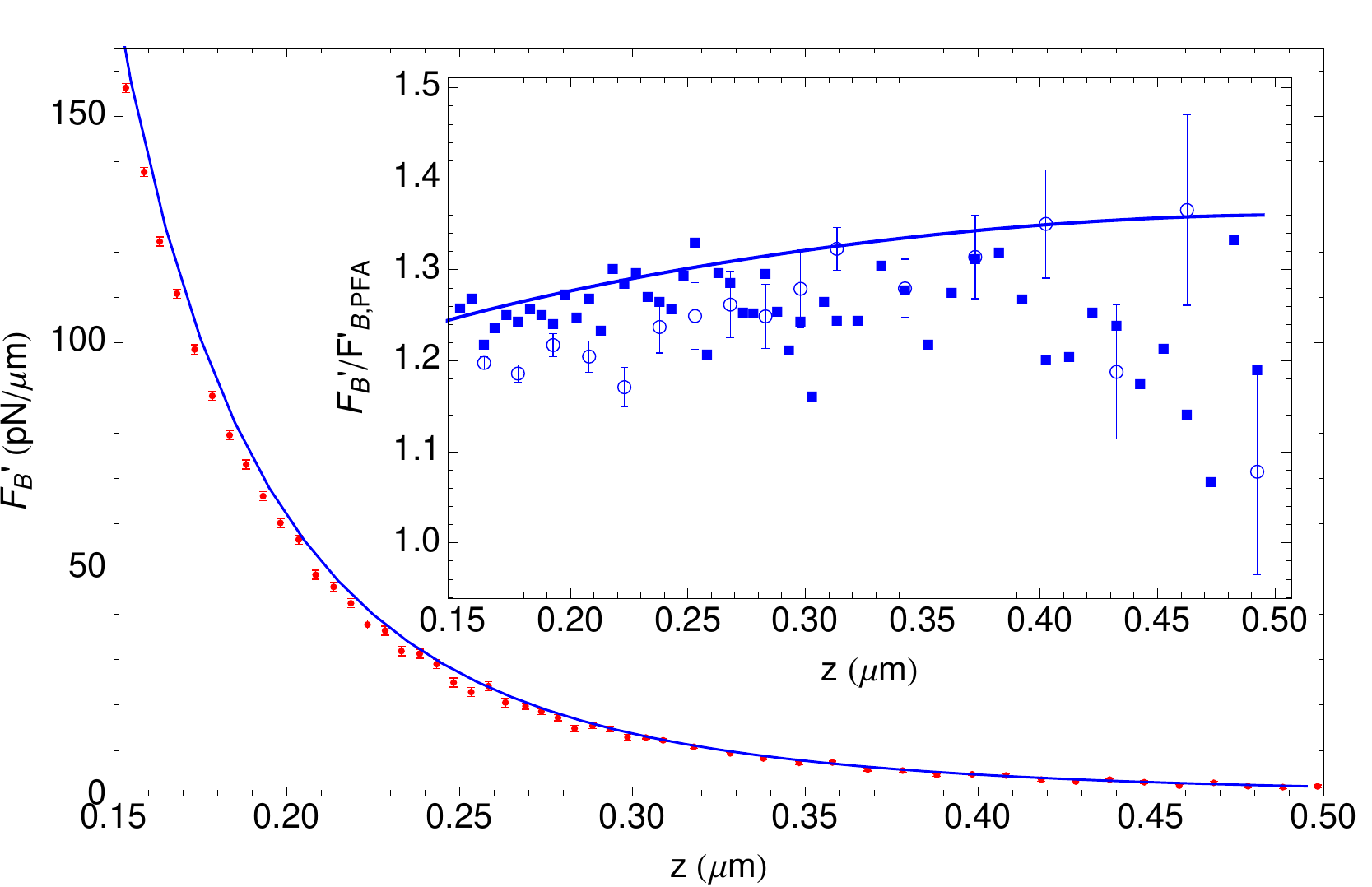}
\caption{Casimir force gradient in the sphere-grating geometry for sample B in \cite{Chan08}: exact numerics (solid) and experimental data (dots with error bars). The inset shows the ratio of the Casimir force gradient divided by the PFA prediction:  exact numerics/theoretical PFA (solid), exact numerics/``experimental" PFA (squares), and experimental data/``experimental'' PFA (hollow circles with error bars).  The geometrical parameters of grating B are: period=400 nm, depth=1070 nm, and filling factor=0.478. The radius of the sphere is $50 \mu$m, and the temperature is set to $T=300$K. }
\label{fig:ForceGradientB}
\end{figure}

We consider the two 1D lamellar gratings used in \cite{Chan08}.
The Casimir force in the plane-grating configuration is evaluated by numerical differentiation of the  plane-grating free energy.
The resulting force gradient in the sphere-gradient geometry calculated with our numerical code is shown in Figs. \ref{fig:ForceGradientA} and \ref{fig:ForceGradientB}, where we also
show the experimental data with their errors, kindly provided to us by H.B. Chan.
Given the uncertainty in the optical parameters used in our numerics compared those of the actual samples, the experiment-theory agreement seems to be very
reasonable. A calculation of the reduced chi square gives $\chi^2_{\rm red}=2.9$ for sample A and $\chi^2_{\rm red}=8.8$ for sample B.
 One can also test the deviations of the exact numerical results from the prediction of the
proximity force approximation (PFA). In this approximation, the force
between the plane and the deeply etched grating is computed by multiplying the force between two plane (i.e., the usual Lifshitz force for the plane-plane geometry) by the filling
factor of the grating (we neglect the contribution of the bottom part of the grating to the PFA result since its depth is sufficiently large). In the insets of Figs. \ref{fig:ForceGradientA} and \ref{fig:ForceGradientB} we plot the ratio between the exact and PFA results for the Casimir force gradients in the sphere-grating configuration. There are three sets of data represented in those insets: a) the solid line is  the ratio of our exact numerics and the theoretical PFA prediction based on Lifshitz theory, both using the parameters above for the optical
data of the samples; b) the hollow circles with their error bars are the ratio of the experimental data of \cite{Chan08} for the force gradient in sphere-grating geometry with respect to the ``experimental" PFA. This last data set was obtained from a separate measurement of the force gradient in the sphere-plane geometry and subsequently multiplied by the grating's filling
factor\cite{Chan08}. Note that since the plane and the grating in \cite{Chan08} were fabricated following identical procedures, one can expect them to have the same optical properties; c)
the squares represent the ratio of our exact numerics and the ``experimental" PFA. Given the uncertainty in the optical parameters used in the numerics, the most unbiased
comparison is to compare case b) against case c). In that case, one is effectively comparing numerators normalized by the same denominator. Again, the theory-experiment
comparison is reasonably good in view of the experimental and theoretical uncertainties.


\begin{figure}
\includegraphics[width=9cm]{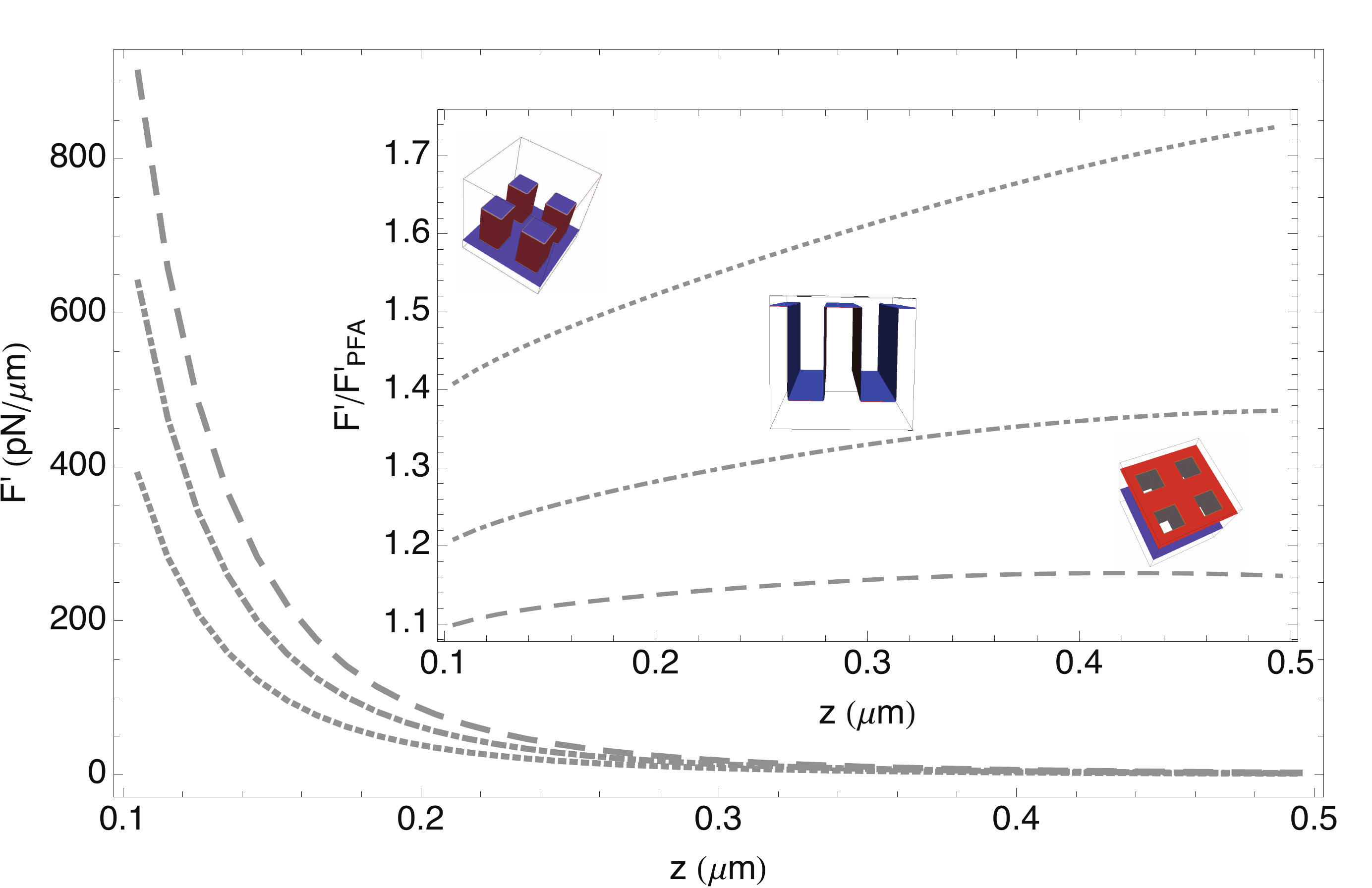}
\caption{Computed Casimir force gradient for 1D and 2D Si gratings. Together with the 1D grating of sample B (dot dashed curve) two simple but representative 2D extensions have been considered. 
The first example is an array of p-doped Si pillars, and  the second  is the complementary structure, i.e. a free standing membrane etched with an array of  square holes. The period of the structures is 400 nm and the filling factor of the pillar and of the membrane are $1/4$ and $3/4$, respectively.  The etch depth for all the structures is 1070 nm. For all cases, a Au sphere with radius $R=50\mu$m has been used. It has also been assumed that the temperature is $T=300$K. We find that the force gradient scales as the  filling factor. The deviation from the PFA is shown in the inset and scales this time with inverse of the filling factor,  having the strongest deviation in the case of the pillars.}
\label{2D-Structures}
\end{figure}

%
%

The numerical errors in the computation of the Casimir force originate from three sources: 1) the truncation of the Matsubara frequency summation, 2) the truncation of the discrete spatial frequency spectrum resulting in finite dimensional reflection matrices, and 3) numerical integration over the continuous transverse wavevector in the first Brillioun zone.  The truncation of the Matsubara summation is determined by the temperature and the minimum distance required in the force displacement curve.  This minimum distance sets the maximum Matsubara frequency for fixed temperature, such that for a minimum spacing of 100 nm at 300K, we find that 36 Matsubara frequencies give free energy convergence of better than 10$^{-4}$.  This frequency cutoff is used in all numerical results presented here. 

The truncation of the discrete spatial frequency (diffraction orders) is necessitated by the need to deal with finite matrices. Our modal approach improves with increasing number of diffraction orders but so does the computational cost. By increasing the number of diffraction orders, we find that the short distance behaviour for the grating is improved.   We have used $N=5$ for all calculations, which was observed to be enough to ensure the convergence at approximately 1\% accuracy. It is important to note that matrices are dense and scale as $4N^4$, which limits practically the spatial frequency cutoff.

The final source of numerical error is the numerical integration over the continuous transverse wavevector in the first Brillioun zone. It is desirable to have minimal $k$-space  sampling to reduce the assembly and computation of the reflection matrices for each Matsubara frequency.   The integration of the continuous wavevector is performed numerically using a an n-point Gauss Legendre quadrature. In order to estimate the integration error, we have computed the the force gradient for 5 different  n-point (9,16,25,36,49) Gaussian quadratures  in the 1st Brillioun zone. The relative error, the standard deviation divided by the mean,  for the experimental displacement range is less than 3\% and quite small at small displacement.  This is quite reasonable since the force gradient at large separations is quite small so small standard deviations give large relative error.  From our three estimates of error,  our  estimated total error in computing the Casimir force is  expected to be less than 3\% over the entire force gradient displacement curve.

The detailed analysis of the 1D lamellar grating problem allowed us to validate the scattering approach by comparing to high precision experimental data.  Here we want to extend the method to 2D periodic structures which provide more tailorable properties in which we can manipulate the Casimir force.  We will consider two simple but representative extensions of the 1D grating. 
The first example is an array of p-doped Si pillars. Similar geometries have been used to obtain a negative index of refraction \cite{Pendry00} in the optical range \cite{Panina02,Grigorenko05}. They have also been exploited to  investigate the phenomenon of quantum reflection \cite{Pasquini04,Pasquini06} of an atom over the purely attractive Casimir-Polder potential generated by the periodic structure. The second case is the complementary structure, i.e. a free standing membrane etched with an array of  square holes. This structure is very similar to the one used to measure the extraordinary light transmission through sub-wavelength apertures \cite{Ebbesen98,Martin-Moreno01,Genet07}.

In Fig.\ref{2D-Structures}  we show the prediction for Casimir force gradient for the sample B 1D grating (dash-dotted line), the pillars (dotted line), and the membrane (dashed line). The period of the structures is in all cases 400 nm and the eteched depth is 1070nm, and the filling factor of the pillars and of the membrane are $1/4$ and $3/4$, respectively. We find that the Casimir force gradient scales as the filling factor, i.e. the force is less attractive for the pillars than for the membrane, with the case of grating in between these two cases. The inset shows the comparison with the relative PFA approximation (filling factor times the Lifshitz force between  plane and sphere). If the PFA were valid, the ratio (inset) would be unity for all separations.  The deviation from the additive approximation scales with the inverse of the filling factor,  having the strongest deviation in the case of the pillars. 




\section{Discussion and Conclusion}

We have developed a  finite-temperature modal approach to compute Casimir interactions between
2D periodic structures. We have compared our computational approach to high precision  published experimental data for 1D lamellar gratings. This benchmark validated our modal approach and led to good agreement between theory and experiment, as confirmed by a reduced chi square values of $2.9$ and $8.8$ for the two samples used in \cite{Chan08}. 

In order to demonstrate the flexibility of our approach, we also calculated the Casimir force between the first simple extensions of a 1D grating, namely an array of square pillars and an array of square holes. In both cases, as already known for the 1D grating, we observe deviation of the Casimir force gradient from the value obtained from the Proximity Force Approximation. This deviation scales with the filling factor and it is more accentuated in the case of the pillars. 

We plan in the near future to extend these results to complex metallic structures, such as 3D metamaterials, which were recognized as possible candidates to engineer the Casimir force between two vacuum separated objects.


\section{Acknowledgments}

We thank A. Contreras-Reyes, R. Depine, S. Johnson, P. Maia Neto, A. McCauley, A. Rodriguez, 
and D. Skigin for helpful comments relating to this research. We are specially thankful to H.B. Chan for
providing us his experimental data and for the ensuing discussions. This work was funded by DARPA/MTO's Casimir Effect Enhancement program under DOE/NNSA Contract
DE-AC52-06NA25396.



\begin{thebibliography}{10}

\bibitem{Casimir48}
H. Casimir, Proc. kon. Ned. Ak. Wet {\bf 51},  793  (1948).

\bibitem{Bordag01}
M. Bordag, U. Mohideen, and V. Mostepanenko, Phys. Rep. {\bf 353},  1  (2001).

\bibitem{Milton04}
K.~A. Milton, J. Phys. A: Math. Gen. {\bf 37},  R209  (2004).

\bibitem{Lamoreaux05}
S.~K. Lamoreaux, Rep. Prog. Phys. {\bf 68},  201  (2005).

\bibitem{Lamoreaux97}
S. Lamoreaux, Phys. Rev. Lett. {\bf 78},  5  (1997).

\bibitem{Mohideen98}
U. Mohideen and A. Roy, Phys. Rev. Lett. {\bf 81},  4549  (1998).

\bibitem{Chan01s}
H. Chan, V. Aksyuk, R. Kleiman, D. Bishop, and F. Capasso, Science {\bf 291},
  1941  (2001).

\bibitem{Bressi02}
G. Bressi, G. Carugno, R. Onofrio, and G. Ruoso, Phys. Rev. Lett. {\bf 88},
  041804  (2002).

\bibitem{Decca03}
R.~S. Decca, D. L\'opez, E. Fischbach, and D.~E. Krause, Phys. Rev. Lett. {\bf
  91},  050402  (2003).

\bibitem{Henkel05}
C. Henkel and K. Joulain, Europhys. Lett. {\bf 72},  929  (2005).

\bibitem{Rosa08a}
F.~S.~S. Rosa, D.~A.~R. Dalvit, and P.~W. Milonni, Phys. Rev. Lett. {\bf 100},
  183602  (2008).

\bibitem{Rahi10a}
S.~J. Rahi, M. Kardar, and T. Emig, Phys. Rev. Lett. {\bf 105},  070404
  (2010).

\bibitem{Silveirinha10}
M.~G. Silveirinha, Phys. Rev. B {\bf 82},  085101  (2010).

\bibitem{Rodrigues07}
R.~B. Rodrigues, P.~A. Maia~Neto, A.~Lambrecht, and S.~Reynaud.
Phys. Rev. A {\bf 75}, 062108 (2007).

\bibitem{Zhao09}
R. Zhao, J. Zhou, T. Koschny, E.~N. Economou, and C.~M. Soukoulis, Phys. Rev.
  Lett. {\bf 103},  103602  (2009).

\bibitem{McCauley10a}
A.~P. McCauley {\it et~al.}, E-print  1006.5489v1  (2010).

\bibitem{Levin10}
M. Levin, A.~P. McCauley, A.~W. Rodriguez, M.~T.~H. Reid, and S.~G. Johnson,
  E-print  arXiv:1003.3487v1  (2010).

\bibitem{Lee01}
S. Lee and W. Sigmund, J. Colloid and Interface Sci. {\bf 243},  365  (2001).

\bibitem{Munday09}
J.~N. Munday, F. Capasso, and V.~A. Parsegian, Nature {\bf 457},  170
  (2009/01/08/print).

\bibitem{Li93a}
L. Li, J. Opt. Soc. Am. A {\bf 10},  2581  (1993).

\bibitem{Petit80}
R. Petit, L. Botten, M. Cadilhac, D. Maystre, P. Vincent, and M. Nevi{\`e}re,
  in {\em Electromagnetic theory of gratings}, edited by R. Petit
  (Springer-Verlag Berlin, New York, 1980).

\bibitem{Neviere98}
M. Neviere and E.~K. Popov,  in {\em Proceedings of SPIE}, edited by W.~R.
  McKinney and C.~A. Palmer (PUBLISHER, ADDRESS, 1998), No.~1, pp.\ 2--10.

\bibitem{Neviere71}
M. Neviere, G. Cerutti-Maori, and M. Cadilhac, Optics Communications {\bf 3},
  48   (1971).

\bibitem{Maystre72}
D. Maystre, Opt. Comm. {\bf 6},  50  (1972).

\bibitem{Johnson11}
S.~G. Johnson,  in {\em Numerical methods for computing Casimir interactions},
  {\em To appear in upcoming Lecture Notes in Physics book on Casimir Physics.}
  (Springer, New York, 2011, E-print: arXiv:1007.0966v1).

\bibitem{Intravaia05}
F. Intravaia and A. Lambrecht, Phys. Rev. Lett. {\bf 94},  110404  (2005).

\bibitem{Intravaia07}
F. Intravaia, C. Henkel, and A. Lambrecht, Phys. Rev. A {\bf 76},  033820
  (2007).

\bibitem{Intravaia09}
F. Intravaia and C. Henkel, Phys. Rev. Lett. {\bf 103},  130405  (2009).

\bibitem{Intravaia10}
F. Intravaia and C. Henkel,  in {\em Proceeding of the Ninth Conference on
  Quantum Field Theory under the Influence of External Conditions}, edited by
  K. Milton and M. Bordag (World Scientific Publishing Co. Pte. Ltd., 5 Toh
  Tuck link, Singapore 596224, 2010), pp.\ 199. E--print: arXiv:0911.3490.

\bibitem{Haakh10}
H. Haakh, F. Intravaia, and C. Henkel, Phys. Rev. A {\bf 82},  012507  (2010).

\bibitem{Chan08}
H.~B. Chan {\it et~al.}, Phys. Rev. Lett. {\bf 101},  030401  (2008).

\bibitem{Lambrecht08a}
A. Lambrecht and V.~N. Marachevsky, Phys. Rev. Lett. {\bf 101},  160403
  (2008).

\bibitem{Lambrecht09}
A. Lambrecht and V.~N. Marachevsky, Int. J. Mod. Phys. A {\bf 24},  1789
  (2009).

\bibitem{Lambrecht11}
A. Lambrecht, A. Canaguier-Durand, R. Gu{\'e}rout, and S. Reynaud,  in {\em
  Casimir effect in the scattering approach: correlations between material
  properties, temperature and geometry}, {\em To appear in upcoming Lecture
  Notes in Physics book on Casimir Physics.} (Springer, New York, 2011,
  E-print: arXiv:1006.2959v2).

\bibitem{Chiu10}
H.-C. Chiu, G.~L. Klimchitskaya, V.~N. Marachevsky, V.~M. Mostepanenko, and U.
  Mohideen, Phys. Rev. B {\bf 81},  115417  (2010).

\bibitem{Gies03}
H. Gies, K. Langfeld, and L. Moyaerts, J. High Energy Phys. {\bf 2003},  018
  (2003).

\bibitem{Bulgac06}
A. Bulgac, P. Magierski, and A. Wirzba, Phys. Rev. D {\bf 73},  025007  (2006).

\bibitem{Emig06}
T. Emig, R.~L. Jaffe, M. Kardar, and A. Scardicchio, Phys. Rev. Lett. {\bf 96},
   080403  (2006).

\bibitem{Rahi09}
S.~J. Rahi, T. Emig, N. Graham, R.~L. Jaffe, and M. Kardar, Phys. Rev. D {\bf
  80},  085021  (2009).

\bibitem{Bordag06}
M. Bordag, J. Phys. A: Math. Gen. {\bf 39},  6173  (2006).

\bibitem{Dalvit06}
D.~A.~R. Dalvit, F.~C. Lombardo, F.~D. Mazzitelli, and R. Onofrio, Phys. Rev. A
  {\bf 74},  020101  (2006).

\bibitem{Lambrecht06}
A. Lambrecht, P.~A.~M. Neto, and S. Reynaud, New J. Phys. {\bf 8},  243
  (2006).

\bibitem{Kenneth08}
O. Kenneth, Phys. Rev. B {\bf 78},  014103  (2008).

\bibitem{Rodriguez07}
A. Rodriguez, M. Ibanescu, D. Iannuzzi, F. Capasso, J.~D. Joannopoulos, and
  S.~G. Johnson, Phys. Rev. Lett. {\bf 99},  080401  (2007).

\bibitem{Rodriguez09}
A.~W. Rodriguez, A.~P. McCauley, J.~D. Joannopoulos, and S.~G. Johnson, Phys.
  Rev. A {\bf 80},  012115  (2009).

\bibitem{Dalibard82}
J. Dalibard, J. Dupont-Roc, and C. Cohen-Tannoudji, J. Phys. France {\bf 43},
  1617  (1982).

\bibitem{Canaguier-Durand09}
A. Canaguier-Durand, P.~A.~M. Neto, I. Cavero-Pelaez, A. Lambrecht, and S.
  Reynaud, Phys. Rev. Lett. {\bf 102},  230404  (2009).

\bibitem{Li97}
L. Li, J. Opt. Soc. Am. A {\bf 14},  2758  (1997).

\bibitem{Ashcroft76}
N.~W. Ashcroft and N.~D. Mermin,  in {\em Solid States Physics}, edited by
  D.~G. Crane (Harcourt College Publisher, New York, 1976).

\bibitem{Kittel96}
C. Kittel, {\em Introduction to Solid State Physics}, 7th ed. (John Wiley and
  Son Inc., New York, 1996).

\bibitem{Jackson75}
J. Jackson, {\em Classical Electrodynamics} (John Wiley and Sons Inc., New
  York, 1975).

\bibitem{Cole68}
R. Cole, {\em Theory of ordinary differential equations}
  (Appleton-Century-Crofts, New York, 1968).

\bibitem{Noponen94}
E. Noponen and J. Turunen, J. Opt. Soc. Am. A {\bf 11},  2494  (1994).

\bibitem{Li96a}
L. Li, J. Opt. Soc. Am. A {\bf 13},  1870  (1996).

\bibitem{Naimark68}
M. Naimark,  in {\em Linear differential operators, Part 1}, edited by W.
  Everitt (New York: Ungar, New York, 1968).

\bibitem{Li96}
L. Li, J. Opt. Soc. Am. A {\bf 13},  1024  (1996).

\bibitem{London35}
F. London and H. London,  in {\em Proceedings of the Royal Society of London},
  No.~866 in {\em Series A, Mathematical and Physical Sciences (1934-1990)}
  (JSTOR, London, 1935), pp.\ 71--88.

\bibitem{Klimchitskaya09c}
G.~L. Klimchitskaya, U. Mohideen, and V.~M. Mostepanenko, Rev. Mod. Phys. {\bf
  81},  1827  (2009).

\bibitem{Canaguier-Durand10}
A. Canaguier-Durand, P.~A.~M. Neto, A. Lambrecht, and S. Reynaud, Phys. Rev.
  Lett. {\bf 104},  040403  (2010).

\bibitem{Canaguier-Durand10a}
A. Canaguier-Durand, P.~A. Maia~Neto, A. Lambrecht, and S. Reynaud, Phys. Rev.
  A {\bf 82},  012511  (2010).

\bibitem{Zandi10}
R. Zandi, T. Emig, and U. Mohideen, Phys. Rev. B {\bf 81},  195423  (2010).

\bibitem{Derjaguin56}
B. Derjaguin, I. Abrikosova, and E. Lifshitz, Quart. Revs. {\bf 10},  33
  (1956).

\bibitem{Bordag10}
M. Bordag and I. Pirozhenko, Phys. Rev. D {\bf 81},  085023  (2010).

\bibitem{Arnold79}
W. Arnold, S. Hunklinger, and K. Dransfeld, Phys. Rev. B {\bf 19},  6049
  (1979).

\bibitem{Bruhl02}
R. Br{\"u}hl {\it et~al.}, Europhys. Lett. {\bf 59},  357  (2002).

\bibitem{Pirozhenko08}
I. Pirozhenko and A. Lambrecht, Phys. Rev. A {\bf 77},  013811  (2008).

\bibitem{Pendry00}
J.~B. Pendry, Phys. Rev. Lett. {\bf 85},  3966  (2000).

\bibitem{Panina02}
L.~V. Panina, A.~N. Grigorenko, and D.~P. Makhnovskiy, Phys. Rev. B {\bf 66},
  155411  (2002).

\bibitem{Grigorenko05}
A.~N. Grigorenko {\it et~al.}, Nature {\bf 438},  335  (2005).

\bibitem{Pasquini04}
T.~A. Pasquini {\it et~al.}, Phys. Rev. Lett. {\bf 93},  223201  (2004).

\bibitem{Pasquini06}
T.~A. Pasquini {\it et~al.}, Phys. Rev. Lett. {\bf 97},  093201  (2006).

\bibitem{Ebbesen98}
T. Ebbesen, H. Lezec, H. Ghaemi, T. Thio, and P. Wolff, Nature {\bf 391},  667
  (1998).

\bibitem{Martin-Moreno01}
L. Martin-Moreno {\it et~al.}, Phys. Rev. Lett. {\bf 86},  1114  (2001).

\bibitem{Genet07}
C. Genet and T.~W. Ebbesen, Nature {\bf 445},  39  (2007).

\end{thebibliography}

%
%
\end{document}